\documentclass{article}

\usepackage{graphicx}
\usepackage{amsthm}
\usepackage{color}
\usepackage{amssymb}

\begin{document}

 \large

\newcommand{\al}{\mbox{$\alpha$}}
\newcommand{\be}{\mbox{$\beta$}}
\newcommand{\ep}{\mbox{$\epsilon$}}
\newcommand{\gam}{\mbox{$\gamma$}}
\newcommand{\sig}{\mbox{$\sigma$}}
\newcommand{\la}{\mbox{$\lambda$}}

\newcommand{\Hil}{\mbox{$\mathbb{H}$}}

\DeclareRobustCommand{\FIN}{%
  \ifmmode % if math mode, assume display: omit penalty etc.
  \else \leavevmode\unskip\penalty9999 \hbox\varphi\nobreak\hfill
  \fi
  $\bullet$ \vspace{5mm}}

\newcommand{\calA}{\mbox{${\cal A}$}}
\newcommand{\calB}{\mbox{${\cal B}$}}
\newcommand{\calC}{\mbox{${\cal C}$}}

\DeclareRobustCommand{\FIN}{%
  \ifmmode % if math mode, assume display: omit penalty etc.
  \else \leavevmode\unskip\penalty9999 \hbox{}\nobreak\hfill
  \fi
  $\bullet$ \vspace{5mm}}

\newcommand{\muas}{\mbox{$\mu$-a.s.}}
\newcommand{\Nat}{\mbox{$\mathbb{N}$}}
\newcommand{\Rea}{\mbox{$\mathbb{R}$}}
\newcommand{\Prob}{\mbox{${I\!\!P}$}}
%%%%%%%%%
%mirar cual era!!!!!!!!!!
\newcommand{\Ze}{\mbox{$\mathbb{Z}$}}
\newcommand{\BBz}{\Ze}
%%%%%%%%%%
\newcommand{\BBe}{\mbox{$\mathbb{E}$}}
\newcommand{\nin}{\mbox{$n \in \mathbb{N}$}}
\newcommand{\suc}{\mbox{$\{X_{n}\}$}}
\newcommand{\sucP}{\mbox{$\{I\!\!P_{n}\}$}}

\newcommand{\conv}{\rightarrow}
\newcommand{\convn}{\rightarrow_{n\rightarrow \infty}}
\newcommand{\convp}{\rightarrow_{\mbox{c.p.}}}
\newcommand{\convs}{\rightarrow_{\mbox{a.s.}}}
\newcommand{\convw}{\rightarrow_w}
\newcommand{\convd}{\stackrel{\cal D}{\rightarrow}}

\newtheorem {Prop}{Proposition} [section]
 \newtheorem {Lemm}[Prop] {Lemma}
 \newtheorem {Theo}[Prop]{Theorem}
 \newtheorem {Coro}[Prop] {Corollary}
 \newtheorem {Claim}[Prop] {Claim}
 \newtheorem {Nota}{Remark}[Prop]
 \newtheorem {Ejem}[Prop] {Example}
 \newtheorem {Defi}[Prop]{Definition}
 \newtheorem {Figu}[Prop]{Figure}
  \newtheorem {Tabla}[Prop]{Table}

\title{A random-projection based procedure to test if a stationary process is Gaussian}%mio:for testing

\author{J.A. Cuesta-Albertos\footnote{Those authors have been  partially supported by the
Spanish Ministerio de Ciencia y Tecnolog\'{\i}a, grant
MTM2005-08519-C02-02 and the Consejer\'{\i}a de Educaci\'on y Cultura de la Junta de Castilla y Le\'on, grant PAPIJCL VA102/06.} $^1$, F. Gamboa$^2$ and A. Nieto-Reyes$^*$ $^1$
\\
[9mm]
$^1$Departamento de
Matem\'{a}ticas, Estad\'{\i}stica y Computaci\'{o}n.
Universidad de Cantabria, Spain
\\
$^2$ Institut de Math\'ematiques de Toulouse, France}
%\date{}
\maketitle

 \begin{abstract}
In this paper we address the statistical problem of testing if a stationary process is Gaussian. The observation consists in a finite sample path of the process. Using a random projection technique introduced and studied in \cite{Cuesta07} in the frame of goodness of fit test for functional data, we perform some decision rules. These rules really stand on the whole distribution of the process and not only on its marginal distribution at a fixed order. The main idea  is to test the Gaussianity on  the marginal distribution of some random linear combinations of the process. This leads to  consistent  decision rules. Some numerical simulations show  the pertinence of our approach.
 \end{abstract}

\vspace{3mm}


\noindent{\em A.M.S. 1980 subject classification: A60G}

\vspace{2.5cm}

\section{Introduction}
\label{intro}
In many concrete situations the statistician observes a finite path $X_1, \ldots, X_n$ of a real temporal phenomena. A common modeling is to assume that the observation is a finite path of a second order weak stationary process ${\mathbf{X}}:=(X_t)_{t\in\BBz}$ (see, for example, \cite{pipo}). This means that the random variable (r.v.) $X_t$ is, for any $t\in\BBz$,  square integrable and that the mean and the covariance structure of the process is invariant  by any translation on the time index. That is, for any $t,s\in\BBz, $ $\BBe(X_t)$ does not depend on $t$ and $\BBe(X_tX_s)$ only depends on the distance between $t$ and $s$. A more popular frame is the Gaussian case where the additional Gaussianity  assumption on all finite marginal distributions of the process  $(X_t)_{t\in\BBz}$ is added. In this case, as the multidimensional Gaussian distribution only depends on moments of order one and two, the process is also strongly stationary. This means that the law of all finite dimensional marginal distributions are invariant if the time is shifted:
$$(X_1,\cdots,X_n)
\stackrel{\mathcal{L}}{=}
(X_{t+1},\cdots,X_{t+n}),\;(t\in\BBz,n\in\Nat).$$
Gaussian stationary process are very popular because they share plenty of very nice properties concerning their statistics or prediction (see, for example, \cite{AzDa} or \cite{Ste}). Hence, an important topic in the field of stationary process is the
implementation of a statistical procedure that allows to assess Gaussianity. In the last three decades, many works have been developed to build such methods.
For example, in \cite{Epps} a test based on  the analysis of the empirical characteristic function is performed. In   \cite{LoVe} based on the skewness and kurtosis test or also called Jarque-Bera test.
In \cite{Moulines} based on both, empirical characteristic function and skewness and kurtosis.
In \cite{subba} we can find another test, this based on the bispectral density function.
An important drawback of these tests is that they only consider a finite order marginal of the process (generally the order one marginal!).  Obviously, this provides tests at the right level for the intended problem; but these tests could be at the nominal power against some non-Gaussian alternatives. For example, in the case of a strictly stationary non-Gaussian process having one-dimensional Gaussian marginal.

In this paper, we propose a procedure to assess that a strictly stationary process  is Gaussian. Our test is consistent against every strictly stationary alternative satisfying regularity assumptions. The procedure  is a combination of the random projection method (see \cite{Cuesta07} and \cite{Cramer}) and  classical methods that allow to assess that the one-dimensional marginal of a stationary process is Gaussian (see the previous discussion).

Regarding the random projection method,  we follow the same methodology as the one proposed in \cite{Cramer}. Roughly speaking, it is shown therein that (only) a random projection characterizes a probability distribution. In particular, we employ the results  of \cite{Cuesta07} where  the main result of \cite{Cramer} is generalized  to obtain goodness-of-fit tests for families of distributions, and in particular for Gaussian families.

Therefore, given a strictly stationary process, $(X_t)_{t \in \mathbb{Z}}$, we are interested in constructing
a test for the null hypothesis
 $H_0: (X_t)_{t \in \mathbb{Z}}$ \mbox{is Gaussian}.
Notice that  $H_0$ holds if, and only if, $(X_t)_{t\leq0}$ is  Gaussian.  So that, using the random projection method, \cite{Cuesta07}, this is, roughly speaking, equivalent to that a (one-dimensional) randomly chosen projection of $(X_t)_{t\leq0}$ is Gaussian.
This idea allows to translate the problem into another one consisting on checking when the one-dimensional marginal of  a random transformation of $(X_t)_{t \in \mathbb{Z}}$ is Gaussian. This can be tested using a usual procedure. Here, we will employ  the well-known Epps test,  \cite{Epps}, and Lobato and Velasco skewness-kurtosis test,  \cite{LoVe}. We also  use a combination of them as a way to alleviate some problems that those tests present.

Furthermore, Epps test checks whether  the characteristic function of the one-dimensional marginal of a strictly stationary process
coincides with the one of a Gaussian distribution. This checking is performed on a fixed finite set of points. As a consequence, it cannot be consistent against every possible non-Gaussian alternative with non-Gaussian marginal. However, in our work, the points employed in Epps test will be also drawn at random. This will provide the consistency of the whole test.
Regarding Lobato and Velasco skewness-kurtosis test we will prove the consistency of the test under different hypothesis than those in \cite{LoVe}.

The paper is organized as follows.
In the next section we will give some basic definitions and notations.
In Section \ref{Juan}, we discuss some useful known results. One concerns the random projection method, some Gaussianity tests for strictly stationary processes and another a procedure for multiple testing. It also contains a new result characterizing Gaussian distributions.
In Section \ref{sectionModelo} we introduce our procedure and analyze its asymptotic behavior. Section  \ref{Simulations} contains some details on the practical application of the method and Section \ref{simu} includes the results of the simulations. The paper ends with a discussion. In the whole paper all the processes are assumed to be integrable.
\section{Notations and basic definitions}
\label{snoba}
If $Y$ is a random variable, we  denote by $\Phi_Y$ its characteristic function; $\Phi_{\mu,\gamma}$ denotes the characteristic function of the Gaussian distribution with mean $\mu\in\Rea$ and variance $\gamma>0$.

 \Hil \ denotes a separable Hilbert space with inner product $\langle \cdot, \cdot\rangle$ and norm $\| \cdot \|$. $\{v_n \}_{n=1}^\infty$  denotes a generic orthonormal
basis of $\Hil$ and $V_n$  the $n$-dimensional subspace spanned by $\{v_1,\ldots,v_n\}$. For any subspace, $V \subset \Hil$ we  write $V^\bot$ for its orthogonal complement. If $\textbf{D}$ is an \Hil -valued random element, then $\textbf{D}_V$  denotes the projection of $\textbf{D}$ on the subspace $V$ of \Hil.

$\mathbf{X}$ and $(X_t)_{t\in\mathbb Z}$ denote indistinctly a process. Through the following, when we say that a process is stationary we mean that it is strictly stationary. Given a stationary process $\mathbf{X},$ let us denote, if they exists, $\mu_X:=\BBe[X_0]$ the mean and $\mu_{X,k}:=\BBe[(X_0-\mu_X)^k],$ with $k\in\mathbb{N},$ the centered moment of order $k.$ Further, let $\gamma_X(t):=\BBe[(X_0-\mu_X)(X_t-\mu_X)],$ with $t\in\mathbb{Z},$ be the autocovariance of order $t$.

Let $X_1, X_2, ..., X_n,$ $n\in\Nat$ be a sample of equally spaced observations of the random process ${\mathbf{X}}.$ Let $\hat{\mu}_X:=n^{-1}\sum_{i=1}^nX_i$ be its sample mean, $\hat{\mu}_{X,k}:=n^{-1}\sum_{i=1}^n(X_i-\hat{\mu}_X)^k,$ for $k\in\mathbb{N},$ its sample centered moment of order $k$ and  $$\hat{\gamma}_X(t):=n^{-1}\sum_{i=1}^{n-|t|}(X_i-\hat{\mu}_X)(X_{i+|t|}-\hat{\mu}_X),$$ for $|t|\leq n-1,$ the sample autocovariance of order $t.$ When it is clear to which process they are referring we suppress the subindex $X.$ Note that then we write $\mu_{X,k}$ as $\mu_k$. For the sake of simplicity, let us denote $\gamma_X :=\gamma_X (0) $ and analogously $\hat{\gamma}_X:=\hat{\gamma}_X(0).$

Finally, by i.i.d.r.vs. we mean independent and identically distributed random variables.

We  assume that all the random elements are defined on the same, rich enough, probability space $(\Omega, \sigma,\Prob)$. %\Prob\mathbb{P}
\section{Preliminary results}
\label{Juan}

In this section we discuss both a characterization of Gaussian distributions in infinite dimensional spaces, a characterization of the one-dimensional Gaussian distributions and two tests of Gaussianity for stationary processes. We also recall some facts on multiple testing procedure. All this material are tools for our results.

%Those are the the building blocks of the test we propose.

Excluding the characterization of the one-dimensional Gaussian distributions (Proposition \ref{Theo1Dimens}), the results in this section are well known and they are included here for the sake of completeness.

\subsection{Characterization of Gaussian distributions in Hilbert spaces}

The result of this subsection comes from \cite{Cuesta07}. It is based on the use of {\it dissipative distributions} which are defined next.

\begin{Defi}
\label{Dede}
Let $\textbf{D}$ be an $\Hil$-valued random element. We will say that its distribution is dissipative if  there exists an orthonormal basis $(v_n)_{n=1}^\infty$ of $\Hil,$  such that
\begin{enumerate}
\item $\Prob \left(\textbf{D}_{V_n^\perp}=0\right)=0$, for all $n\geq 2$ (see Section \ref{snoba} for the definition of $V_n$).
\item The conditional distribution of $\textbf{D}_{V_n}$ given $\textbf{D}_{V_n^\perp}$ is absolutely continuous with respect to the $n$-dimensional Lebesgue measure.
\end{enumerate}
%
%Here, $V_n$ denotes the linear space generated by $v_1,\cdots,v_n$ $(n\geq 1)$.
\end{Defi}

Theorem 3.6 in \cite{Cuesta07} states the following:

\begin{Theo}[Cuesta-Albertos et al. (2007)] \label{TheoGausianas}
 Let $\eta$ be a dissipative distribution on $\Hil$. If $\textbf{X}$ is  an \Hil -valued random element and
$$
\eta(\{\textbf{h}\in\Hil:\, \mbox{ the distribution of } \langle \textbf{X},\textbf{h} \rangle \mbox{ is Gaussian} \})>0,
$$
 then $\textbf{X} $ is Gaussian.
\end{Theo}

The importance of this result relies on the fact that if $\eta$ is dissipative then the following $0-1$ law  holds
\[%begin{equation} \label{EqUnid1}
\eta (\left \{\textbf{h}\in\Hil :  \mbox{ the distribution of } \langle \textbf{X},\textbf{h} \rangle \mbox{ is Gaussian}   \right\} )\in \{0,1\}.
\]%end{equation}
Moreover, $\textbf{X}$ is not Gaussian if, and only if,
\[%begin{equation} \label{EqUnid1}
\eta(\left \{\textbf{h}\in\Hil :  \mbox{ the distribution of } \langle \textbf{X},\textbf{h} \rangle \mbox{ is Gaussian} \right\})=0.
\]%end{equation}

In other words, if we ask if the distribution of $\textbf{X}$ is Gaussian, then  the only thing we have to do is to select at random a point $\textbf{h}\in\Hil$ using a dissipative distribution  and check if the real-valued random variable  $\langle \textbf{X},\textbf{h} \rangle$ is Gaussian. We will obtain the right answer with probability one.

\subsection{Characterization of  one-dimensional Gaussian distributions}
We start this subsection by stating the definition of analytic characteristic function which has been taken  from \cite{Laha}.

\begin{Defi}
A characteristic function $\Phi$ is said to be analytic  if there exist
\begin{itemize}
\item
 a complex valued function, $\phi$, of the complex variable $z$ which is holomorphic in a circle $\{z:|z|<\rho\},$ where $\rho >0,$

  \item a positive real number $\delta$ such that $\Phi(t)=\phi(t),$ for $|t|<\delta.$

 \end{itemize}
\end{Defi}

That is, an analytic characteristic function is a characteristic function which coincides with a holomorphic function in some neighborhood of zero.

Some properties on analytic characteristic functions may be found in \cite{Laha}. In particular,  it is proved therein that the characteristic function of a Gaussian distribution is analytic (this is a well known fact). Some other well-known distributions having analytic characteristic function are the binomial, Poisson and gamma distributions but not the Cauchy one.

The following result will be useful to assess that  our goodness of fit test will work with all non-Gaussian alternatives.%is universal.

\begin{Prop}\label{Theo1Dimens}
Let $P$ be a Borel probability measure defined on \Rea. Assume that $P$ is absolutely continuous with respect to the Lebesgue measure. Let $Y$ be a r.v.  having an analytic characteristic function $\Phi_Y$.

Then, $Y$ is Gaussian if, and only if,
\begin{equation} \label{EqUnid1}
\exists m \in \Rea,\; \exists s\in\Rea^+ \mbox{ such that } P (\left \{y \in \Rea: \Phi_Y(y)= \Phi_{m, s}(y) \right \} )>0.
\end{equation}

\end{Prop}
\begin{proof}
\ \\
Necessary part is obvious. Let us show the sufficiency. As $Y$ satisfies (\ref{EqUnid1}), and
$P$  is absolutely continuous, we have that the set
$R:=\{y\in\Rea : \Phi_Y(y)= \Phi_{m, s}(y)  \}$ is infinite and not denumerable.
Thus, it contains at least an accumulation point.

Furthermore, the function $y\rightarrow  \Phi_Y(y)-\Phi_{m, s}(y)$ is analytic,
and it vanishes on $R$.
%$\{y\in\Rea : \Phi_Y(y)= \Phi_{m, s}(y)  \}.$
 Therefore, this function has a non-isolated
zero but the only analytical function with at least a non-isolated zero is the null function which
proves the result (see for example \cite{Rud}).
%\qed
\end{proof}

Proposition \ref{Theo1Dimens} may be seen as a spectral counterpart of
%to check whether a given one-dimensional variable is Gaussian or not is similar to the use of
Theorem \ref{TheoGausianas}.
% to check whether an infinite-dimensional vector is Gaussian.

\subsection{Classical tests of Gaussianity for stationary processes}
\label{SubsPro}
Through this section we present some useful popular tests for checking whether a  stationary random process $(Y_t)_{t \in \mathbb{Z}},$ is Gaussian.
%As we said in the introduction,   $Y_1, Y_2, ..., Y_n,$ $n\in\Nat$ is a sample of equally spaced observations of the process.

\subsubsection{Epps test} \label{SubsEpps}

The test discussed in this section is a particular case of the one studied in Section 3 of \cite{Epps}.
 We begin with  some notations and definitions. Given $N>1$,  let us define
$$
\Lambda_N:=\{\lambda:=(\lambda_1  ,\ldots , \lambda_N)^T\in\Rea^+_N: \lambda_i\neq \la_j \mbox{ for all } i\neq j, \mbox{ } i,j=1,...N\},
$$
where $^T$ denotes transposition.\\
Let $\lambda \in \Lambda_N$ and let $\hat g(\lambda)$ be  the $2N$-dimensional column vector composed by the real and complex parts of the empirical characteristic function computed at  $\lambda$. That is
\[
\hat g(\lambda) := \frac 1 n \sum_{i=1}^n (\cos(\lambda_1 Y_i), \sin(\lambda_1 Y_i),\ldots ,\cos(\lambda_N Y_i),\sin(\lambda_N Y_i))^T.
\]

Further let,  for $\nu\in\Rea$ real and $\rho>0$, the $2N$-dimensional vector composed by the real and complex parts of $\Phi_{\nu ,\rho }$ computed at  $\lambda$:
\[
g_{\nu ,\rho}(\lambda) :=
\left( \mbox{Re} (\Phi_{\nu ,\rho}(\lambda_1)), \mbox{Im} (\Phi_{\nu ,\rho}(\lambda_1)), \ldots , \mbox{Re} (\Phi_{\nu ,\rho }(\lambda_N)), \mbox{Im} (\Phi_{\nu ,\rho }(\lambda_N))\right)^T.
\]
We denote by $f_{\mathbf{Y}}(0 ,(\mu_Y,\gamma_Y), \lambda)$ the spectral density matrix (see for example \cite{And}) of the process
$$
\left(g(Y_t,\lambda)\right)_{t \in \mathbb{Z}}:=\left((\cos(\lambda_1 Y_t),\sin(\lambda_1 Y_t),\ldots ,\cos(\lambda_N Y_t),\sin(\lambda_N Y_t))\right)^T_{t \in \mathbb{Z}}
$$
at frequency $0$.
Notice that if we assume that $(Y_t)_{t \in \mathbb{Z}}$ is a Gaussian stationary process  with
\begin{equation}\label{H}
\sum_{t \in \mathbb{Z}} |t|^\zeta| \gamma_{\mathbf{Y}} (t) | < \infty, \mbox{ for some } \zeta>0, \end{equation}
then the existence of $f_{\mathbf{Y}}(0 ,(\mu_Y,\gamma_Y), \lambda)$ is one of the conclusions of Lemma 2.1 in \cite{Epps}.
For the construction of the test statistic we will use the following estimator of $f_{\mathbf{Y}}(0 ,(\mu_Y,\gamma_Y), \lambda)$:
\begin{eqnarray}\label{fhat} \hat f(0,\lambda)=(2\pi n)^{-1}
 \left(
 \sum_{t=1}^{n}\hat G(Y_{t,0} ,\lambda)+2\sum_{i=1}^{\lfloor n^{2/5}\rfloor}(1-i/\lfloor n^{2/5}\rfloor)
 \sum_{t=1}^{n-i}\hat G(Y_{t,i},\lambda)
 \right),\end{eqnarray}
where $\hat G(Y_{t,i},\lambda)=(g(Y_t,\lambda)-\hat g(\lambda))(g(Y_{t+i},\lambda)-\hat g(\lambda))^T$ and $\lfloor \cdot \rfloor$ denotes the integer part.  The estimator (\ref{fhat}) was used in \cite{Epps}, but with $2/5$ replaced by a general constant in the interval $(0,1/2).$ Notice also that it is a particular case of the one proposed in \cite{Gapos}. In \cite{Epps} it is proved that if $(Y_t)_{t \in \mathbb{Z}}$ is Gaussian, stationary and satisfies (\ref{H}), then $ \hat f(0,\lambda)$
 converges almost surely to $f_{{\mathbf{Y}}}(0 ,(\mu_Y,\gamma_Y), \lambda).$
Let $G^+_n(\lambda)$ be the generalized inverse of $2 \pi \hat f(0,\lambda)$ and let $Q_n(\nu ,\rho,\lambda)$ be the quadratic form
\begin{equation} \label{EqFCuadrat}
Q_n(\nu ,\rho,\lambda):=
\left(\hat g(\lambda) -  g_{\nu ,\rho}(\lambda)\right)^T
G^+_n(\lambda)
\left(\hat g(\lambda) -  g_{\nu ,\rho}(\lambda)\right).
\end{equation}

Let $\Theta$ be an open bounded subset of $\Rea\times\Rea^+$ and let $\la\in\Lambda_N$. We state two assumptions.
\begin{description}
\item[H1.] The set $\Theta_0(\la):=\{(\nu,\rho)\in\Theta:\Phi_{\nu,\rho}(\la_i)=\Phi_{\mu_Y,\gamma_Y}(\la_i), i=1,...,N\}$ is nowhere dense in $\Theta.$
\item[H2.] For each $(\nu,\rho)\in\Theta_0(\la)$ we have, $f_{{\mathbf{Y}}}(0,(\nu,\rho),\la)=f_{{\mathbf{Y}}}(0,(\mu_Y,\gamma_Y),\la)$  and  $$\left.\frac{\partial \Phi_{x,y}(\la_i)}{ \partial (x,y)}\right|_{(x,y)=(\nu,\rho)}=\left.\frac{\partial \Phi_{x,y}(\la_i)}{ \partial (x,y)}\right|_{(x,y)=(\mu_Y,\gamma_Y)}, \mbox{ } i=1,...,N.$$
\end{description}

Theorem \ref{TheoEpps} below describes  the Gaussianity test studied in \cite{Epps}.
\begin{Theo}[Epps (1987)]\label{TheoEpps}
Let $(Y_t)_{t \in \mathbb{Z}}$  be a  stationary Gaussian process satisfying (\ref{H}). Let $\Theta$ be an open and bounded subset of $\Rea\times\Rea^+$ and $\la\in\Lambda_N$ such that \emph{\textbf{H1.}} and \emph{\textbf{H2.}} hold.  Further, let  $(\mu_n ,\gamma_n)$ be the minimizer on $\Theta$ nearest to $(\hat\mu_Y,\hat\gamma_Y)$ of the map
\[
(\nu ,\rho) \conv Q_n(\nu ,\rho,\lambda).
\]
 Assume further that $f_{{\mathbf{Y}}}(0 ,(\mu_Y,\gamma_Y), \lambda)$ is positive definite.
Then, for each fixed $\lambda\in\Lambda_N$, $nQ_n(\mu_n ,\gamma_n,\lambda)$ converges in distribution to $\chi^2_{2N-2}$.
%a central chi-squared distribution with $2(N-1)$ degrees of freedom.
\end{Theo}

\begin{Nota} {\rm Obviously a test based on Theorem \ref{TheoEpps}  may be not consistent. Indeed, it only focuses on the values of the characteristic function at some points.
%of the random variables $Y_t$  at the points given by the marginals of the fixed $\lambda$.
In other words, the test could not detect some alternatives with Gaussian one-dimensional marginal. Even the test fails against alternatives with non-Gaussian one-dimensional marginal but that satisfy that the characteristic functions of the one-dimensional marginal coincides with the one of the corresponding Gaussian at the selected points.}
\end{Nota}

\subsubsection{Lobato and Velasco test}\label{LoVeprevio}
The test to assess normality of time series discussed in this Subsection was introduced in \cite{LoVe}. It uses the skewness-kurtosis test statistic, also called Jarque-Bera test (see \cite{Bo} and \cite{Jarque}), but improves previous tests of this kind because the statistic is studentized by standard error estimators.

Given a process $\bf Y,$ let us denote $\tilde{F}_k:=2\sum_{t=1}^{n-1}\hat{\gamma}_Y(t)(\hat{\gamma}_Y(t)+\hat{\gamma}_Y(n-t))^{k-1}+\hat{\gamma}_Y^k$.  This is  an estimator of $F_k:=\sum_{t=-\infty}^\infty\gamma_Y(t)^k.$
The test  proposed in \cite{LoVe} handles the statistic: \[\tilde G_Y=\frac{n\hat{\mu}_{Y,3}^2}{6\tilde{F}_3}+\frac{n(\hat{\mu}_{Y,4}-3\hat{\mu}_{Y,2}^2)^2}{24\tilde{F}_4}. \]

\begin{Theo}[Lobato and Velasco (2004)]\label{LoVe}
Let $(Y_t)_{t \in \mathbb{Z}}$ be an ergodic stationary  process.
\begin{itemize}
\item \label{11} If $(Y_t)_{t \in \mathbb{Z}}$ is Gaussian and satisfies $\sum_{t=0}^\infty|\gamma_{\bf Y}(t)|<\infty,$  then
$\tilde{G}_Y \longrightarrow\chi_2^2$ in distribution.
\item  \label{nada} If $(Y_t)_{t \in \mathbb{Z}}$  satisfies
\begin{itemize}\item $\BBe[Y_t^{16}]<\infty$,
\item
$\sum_{t_1=\infty}^\infty\cdot\cdot\cdot\sum_{t_{q-1}=\infty}^\infty|k_q(t_1,...,t_{q-1})|<\infty,$  for q=2,...,16, where $k_q(t_1,...,t_{q-1})$ denotes the $q$th-order cumulant of $Y_1, Y_{1+t_1},...,Y_{1+t_{q-1}},$
\item  $\sum_{t=1}^\infty[\BBe|(\BBe(Y_0-\mu)^k|{\cal F}_{-t})-\mu_k|^2]^{1/2}<\infty,$  for $k=3,4,$ where ${\cal F}_{-t}$ denotes the $\sigma$-field generated by $Y_j$, $j\leq-t,$ and
    \item $\BBe[(Y_0-\mu)^k-\mu_k]^2+2\sum_{t_1=\infty}^\infty \BBe([(Y_0-\mu)^k-\mu_k][(Y_t-\mu)^k-\mu_k])>0,$  for $k=3,4,$
\end{itemize}
then the statistic $\tilde{G}_Y$ diverges to infinity whenever $\mu_{Y,3}\neq 0$ or $\mu_{Y,4}\neq 3\mu_{Y,2}^2.$
\end{itemize}
\end{Theo}
In  Section \ref{sectionModelo} we will prove this theorem under lighter assumptions on the alternative. We will need the following recent result taken from \cite{Kava}. It is an improvement of the well-known result in \cite{An}.

\begin{Theo}[Kavalieris (2008)]\label{teoKava}
Let  $(Y_t)_{t \in \mathbb{Z}}$ be a stationary process with the representation
\begin{eqnarray} \label{Kava}
Y_t=\sum_{i=1}^\infty k(i)\epsilon_{t-i}, \sum_{i=1}^\infty |k(i)|<\infty, \sum_{i=1}^\infty ik(i)<\infty,  \BBe[\epsilon_n]=0,
\mbox{where }  (\epsilon_t) \mbox{ are i.i.d.r.vs..}\end{eqnarray}  Assume that $\BBe[|\epsilon_n|^\alpha]<\infty$ for some $2<\alpha<4.$ If $\tau_n<cn^\beta$ for $0<\beta<1$ and  $c>0,$ then
$$\max_{0\leq t\leq \tau_n}|\hat\gamma(t)-\gamma(t)|=o(n^{2/\alpha-1}) \mbox{ almost surely}.$$
\end{Theo}

\subsection{Multiple testing}\label{MT}
In Section \ref{Simulations} we will propose to use several tests to assess the Gaussianity of a process. Thus we obtain several $p$-values $p_1,\ldots, p_k$, where $k$ is the number of procedures used.

The most popular way to handle several $p$-values is to use the Bonferroni correction. However, it is very well-known that this procedure is too conservative. Several alternatives have been proposed in the literature in order to alleviate this problem. Here, we will employ the  {\it false discovery rate} (FDR). The FDR is the expected proportion of wrongly rejected hypotheses along the $k$ tests. Taking into account that all the hypothesis we have are equivalent, the FDR coincides with the level of the procedure.

The FDR was introduced in Benjamini and Hochberg \cite{BH} for independent tests. Here, we employ the improvement proposed  in \cite{Yekutieli} that does not require dependence assumptions among the tests. This procedure, when applied to our case, works as follows:

\begin{Theo}[Benjamini and Yekutieli (2001)] \label{TheoFDR}
Let us assume that we apply $k$ statistical tests to check the same null hypothesis and that  the ordered $p$-values that we obtain are $p_{(1)}, ...,p_{(k)},$ where $p_{(1)}\leq  \ldots \leq p_{(k)}$.

Let $\alpha \in (0,1)$. The FDR of the test which rejects if the set
\[
\left\{i:p_{(i)}\leq \frac{i \alpha}{k\sum_{j=1}^k j^{-1}} \right\}
\]
is not empty is, at most, $\alpha$.
\end{Theo}
Therefore, according to the previous theorem, if we denote $$p_0=k\sum_{j=1}^k j^{-1}\min_{i=1,...,k}p_{(i)}/i$$ we can reject at any level $\alpha\geq p_0$ and then, we can take $p_0$ as the resulting $p$-value of the procedure.

%%%%%%%%%%%%%%%%%%%%%%%%%%%%%%%%%%%%%%%%%%
%%%%%%%%%%%%%%%%%%%%%%%%%%%%%%%%%%%%%%%%%%%
%%%%%%%%%%%%%%%%%%%%%%%%%%%%%%%%%%%%%%%%%%%%
\section{A Gaussianity test for stationary processes}\label{sectionModelo}
In this section we present a universal test to check if the distribution of a  stationary process  is Gaussian. Thus, given $\textbf{X}:=(X_t)_{t\in\mathbb{Z}}$ a stationary
process of real-valued random variables  we are interested in constructing a test for the null hypothesis
\begin{itemize}
\item[  ] $H_0: \textbf{X} \mbox{ is Gaussian}$
\end{itemize}
against the alternative
\begin{itemize}
\item[  ] $H_a: \textbf{X} \mbox{ is not Gaussian.} $
\end{itemize}

Notice  $H_0$ holds if, and only if, for all $t\in\mathbb{N},$ $(X_1,\ldots, X_t)^T$ is a Gaussian vector. As \textbf{X} is stationary, it is equivalent to  the distribution of $(X_t)_{t\leq0}$ is Gaussian. In addition, it is the sames as the Gaussianity of the process $X^{(t)}:=(X_j)_{j\leq t},$ for any $t\in\mathbb{Z}.$
%infinite-dimensional vector $( \ldots,X_0)^T$ is Gaussian; is equivalent to the Gaussianity of the process $X^{(t)}:=( \ldots,X_t)^T,$ for any $t\in\mathbb{Z}.$
To check whether $X^{(t)}$ is Gaussian, we only need to include $X^{(t)}$  in an appropriate Hilbert space, then select a vector $\textbf h$ using a dissipative distribution, and compute the scalar product $\langle X^{(t)},\textbf h \rangle$ because, according to Theorem \ref{TheoGausianas}, almost surely, $X^{(t)}$ is Gaussian if, and only if, $\langle X^{(t)},\textbf h \rangle$ is Gaussian.

Concerning the Hilbert space in which the  process is included, let us consider the space of sequences
$$l^2 = \left\{(x_n)_{n\in\mathbb{N}}: \sum_{n\in\mathbb{N}} x_n^2 a_n < \infty \right\} ,$$
with $a_0:=1$ and $a_n=n^{-2}, (n\geq 1)$  endowed with the scalar product
\[\langle \textbf  x, \textbf y \rangle
=
\sum_{n\in\mathbb{N}} x_ny_n a_n, \mbox{ where } \textbf  x = (x_n)_{n\in\mathbb{N}}, \textbf y = (y_n)_{n\in\mathbb{N}}.\]

It is easy to see that if $\textbf{X}$ is a stationary process and if the variance of $X_t$ is finite, then, almost surely, $X^{(t)} \in l^2 $ and that the Gaussianity in this space is equivalent to the (usual sense) Gaussianity of $X^{(t)}.$ The reason is that $E[\sum_{n\in\Nat}X^2_{t-n}a_n]$ is finite if it is so the variance of $X_t.$

Now we need a dissipative distribution on $l^2.$ We will use the so-called Dirichlet distribution (see \cite{Pit}).
We build this distribution using the so-called stick breaking method.
That is, let $\alpha_1,\alpha_2>0$  and consider the probability distribution which selects a random point in $l^2$ according to the following iterative procedure:

\begin{itemize}

\item
$l_0\in [0,1]$ is chosen with the beta distribution of parameters $\alpha_1$ and  $\alpha_2.$

\item
Given $n\geq 1$, $l_{n}\in [0,1-\sum_{i=0}^{n-1}l_i]$ is chosen with the beta distribution of parameters $\alpha_1$ and  $\alpha_2,$ times $1-\sum_{i=0}^{n-1}l_i.$

\end{itemize}

Let us define $H_n=(l_n /a_n)^{1/2}$ for $n\geq0$ and take $\textbf H=(H_0,...)^T.$ It can be easily checked that the distribution of $\textbf H$ is  dissipative (see Definition \ref{Dede}). Moreover, $\textbf H\in l^2$ almost surely because, as shown in  Proposition \ref{unop}, $\| {\bf H}\|=1$, almost surely.

\begin{Prop} \label{unop}
Let ${\bf H} = (H_n)_{n \geq 0}$  be a stochastic process constructed as described above. Let $\alpha:=\alpha_1/(\alpha_1+\alpha_2)$ be the mean of the beta distribution of parameters $\alpha_1$ and  $\alpha_2.$ Then, we have that

\begin{enumerate}
\item \label{El}
$\BBe[l_n]=\alpha(1-\alpha)^n, \mbox{ for every } n\in\Nat^*$.
\item  \label{El.2}
$\| {\bf H }\|= 1$, almost surely.
\end{enumerate}
\end{Prop}
\begin{proof}
\ \\
Obviously {\it \ref{El}.} holds for $n=0$. Thus, let us assume that {\it \ref{El}.}  is satisfied for $n \in \Nat^*$ and let us show that it also holds for $n+1$. By construction,  we have that if $\beta$ is a random variable with beta distribution of parameters $\alpha_1$ and  $\alpha_2$, then
\[
\BBe[l_{n+1}] = \BBe[\beta] \left( 1 - \sum_{i=0}^n \BBe[l_{i}] \right)
= \alpha \left( 1 - \sum_{i=0}^n  \alpha(1-\alpha)^i \right)
=
  \alpha(1-\alpha)^{n+1} ,
\]
where last equality comes from the application of the formula giving the sum of $n$ numbers in a geometric progression.
Concerning {\it \ref{El.2}.}, we have that
\begin{equation}\label{EqEl1}
\| {\bf H }\| = \sum_{i=0}^\infty H_i^2 a_i= \sum_{i=0}^\infty l_i \leq 1.
\end{equation}
Indeed,  by construction, for every \nin , $\sum_{i=0}^n l_i \leq 1$. However, applying {\it \ref{El}.}, we have that
\[
\BBe\left[ \| {\bf H }\|\right] =  \sum_{i=0}^\infty   \alpha(1-\alpha)^{i}= 1.
\]
So that, by (\ref{EqEl1}) we obtain \textit{\ref{El.2}.}
%\qed
\end{proof}

Now, let $\textbf h=(h_i)_{i\in \mathbb{N}}$ be a fixed realization of the random element $\textbf{H},$ drawn independently of the process $\textbf X$.
%and which remains fixed through the entire Section \ref{sectionModelo}.
Let us  consider the process $\textbf{Y}=(Y_t)_{t \in \mathbb{Z}}$  given by the projections of
%the process
$(X^{(t)})_{t\in\mathbb{Z}}$ on the one-dimensional subspace generated by $\textbf{h},$ i.e.
\begin{eqnarray}\label{Y}
Y_t=\sum_{i=0}^{\infty} h_i
X_{t-i}a_i, t \in \mathbb{Z}.
\end{eqnarray}

As we will see in  Proposition \ref{prop}, the properties of the process  \textbf X are inherited by the process  \textbf Y.  Moreover, according to  Theorem \ref{TheoGausianas}, to assess the Gaussianity of \textbf X is enough to do the job on the one-dimensional  marginal distributions of \textbf Y.  This can be done for instance with Epps or Lobato and Velasco tests presented in Section \ref{SubsPro} whenever $\textbf Y$ satisfies the appropriate  assumptions. The following Subsections are devoted to this task.

We begin by proving Lemma \ref{LL} which is necessary for Proposition \ref{prop}.

\begin{Lemm}\label{LL}
Let $\textbf{X}$ be an ergodic and stationary process such that $\sum_{t=0}^\infty|\gamma_X(t)|<\infty.$
If we select $\textbf{H}$ as described above, then,
\begin{enumerate}
\item \label{un}$\sum_{i=0}^\infty H_ia_i<\infty$ almost surely.
\item \label{do} Almost surely, the random variable $L:=\sum_{i,j=0}^\infty H_i H_j a_i a_j |X_{-i}-\mu_X| |X_{t-j}-\mu_X|$ is conditionally integrable given $\bf H$.
\end{enumerate}
\end{Lemm}
\begin{proof}
\ \\
\textit{\ref{un}.} It is straightforward because  the  Cauchy-Schwartz inequality  gives that
$$
\sum_{i=0}^\infty H_ia_i\leq\left( \sum_{i=0}^\infty  l_i\right)^{1/2} \left( 1+ \sum_{i=1}^\infty  1/ {i^2} \right)^{1/2}=\left( 1+ \sum_{i=1}^\infty  1/ {i^2} \right)^{1/2}<\infty,
\mbox{ almost surely,}
$$
where last equality comes from Proposition \ref{unop}.\\
To prove \textit{\ref{do}.}, let  ${\bf h}=(h_0,h_1,\ldots)$ be a fixed realization of $\bf H$.  We have that
\begin{eqnarray*}
\BBe[L|{\bf h}]
& =&
\sum_{i,j=0}^\infty h_ih_ja_ia_j\BBe[|X_{-i}-\mu_X||X_{t-j}-\mu_X|]
\\
&\leq&
 \sum_{i,j=0}^\infty h_ih_ja_ia_j(\BBe[(X_{-i}-\mu_X)^2])^{1/2}(\BBe[(X_{t-j}-\mu_X)^2])^{1/2}=\gamma_X\left(\sum_{i}^\infty h_ia_i\right)^2.
\end{eqnarray*}
 Thus, $L$ is conditionally integrable thanks to \textit{\ref{un}.} and that  $\gamma_X\leq \sum_{t=0}^\infty |\gamma_X(t)|<\infty$.
%
%\qed
\end{proof}
In the sequel
 $\gamma_{Y|{\bf h} } (t)$   denotes the conditional autocovariance of order $t$ of $Y$ given $\bf h$. That is,  denoting by $\mu_{Y|{\bf h}}$ the conditional expectation of $Y_0$ given  $\bf h $,
$$\gamma_{Y|{\bf h} } (t):=\BBe[(Y_0-\mu_{Y|{\bf h}})(Y_t-\mu_{Y|{\bf h}})|{\bf h}].$$

\begin{Prop}\label{prop}
Let $(X_t)_{t\in\mathbb{Z}}$ be an ergodic and stationary process such that $\sum_{t=0}^\infty t^\zeta|\gamma_X(t)|<\infty,$ with $\zeta\geq 0.$
Then, conditionally on $\bf h$, the process $(Y_t)_{t\in\mathbb{Z}}$ defined  in (\ref{Y}) is ergodic and  stationary. In addition,  $\BBe[|Y_0| | \bf{h} ]$ and $\sum_{t=0}^\infty t^\zeta|\gamma_{Y|{\bf h} }(t)|$ are finite.
\end{Prop}
\begin{proof}\ \\
 $(X_t)_{t\in\mathbb{Z}}$ is a stationary ergodic process. So that, conditionally on $\bf h$, $(Y_t)_{t\in\mathbb{Z}}$ is also a stationary ergodic  process (see \cite{Doob} page 458).

Using the definition of the process \textbf Y  we have
$$
\BBe[|Y_0| \ | \textbf{h} ] \leq \BBe\left. \left[\sum_{i=0}^\infty h_ia_i |X_{-i}| \right| \textbf{h} \right]=\BBe[|X_0|]\sum_{i=0}^\infty h_ia_i<\infty , \ \mbox{ a.s.}
$$
because  of \textit{1.} in  Lemma \ref{LL}.% and $\BBe[|X_0|]<\infty$.

By \textit{\ref{do}.} in Lemma \ref{LL}, we have that   $$\gamma_{Y|{\bf h}}(t)=\BBe[\sum_{i,j=0}^\infty h_ih_ja_ia_j(X_{-i}-\mu_X)(X_{t-j}-\mu_X)\ | {\bf h}]$$ exists. So that, using the dominated convergence theorem, we obtain that $$\gamma_{Y|{\bf h}} (t)=\sum_{i,j=0}^\infty h_ih_ja_ia_j\gamma_X(t-j+i)$$ and
$$\sum_{t=0}^\infty t^\zeta|\gamma_{Y|{\bf h}}(t)|\leq \sum_{i,j=0}^\infty h_ih_ja_ia_j\sum_{t=0}^\infty t^\zeta|\gamma_X(t-j+i)|.$$
Obviously, $\sum_{i,j=0}^\infty h_ih_ja_ia_j\sum_{t=0}^\infty t^\zeta|\gamma_X(t-j+i)|=:T_1+T_2+T_3$, where
\begin{itemize}
\item $T_1=\sum_{j=0}^\infty h_ja_j \sum_{i=j}^\infty h_ia_i \sum_{t=0}^\infty
t^\zeta|\gamma_X(t-j+i)|,$
\item $T_2= \sum_{j=0}^\infty h_ja_j\sum_{i=0}^{j-1} h_ia_i \sum_{t=2j+1}^\infty
t^\zeta|\gamma_X(t-j+i)|,$
\item $T_3=\sum_{j=0}^\infty h_ja_j\sum_{i=0}^{j-1}h_ia_i \sum_{t=0}^{2j}t^\zeta|\gamma_X(t-j+i)|.$
\end{itemize}
If $i \geq j,$ as  $t\in\Nat^*$ and
$\zeta\geq 0$, we have $t^\zeta\leq(t-j+i)^\zeta.$ Thus,
$$
T_1\leq
\sum_{j=0}^\infty h_ja_j \sum_{i=j}^\infty h_ia_i \sum_{t=0}^\infty
(t-j+i)^\zeta|\gamma_X(t-j+i)|\leq \sum_{j=0}^\infty h_ja_j \sum_{i=j}^\infty
h_i a_i\sum_{t=0}^\infty t^\zeta|\gamma_X(t)|,
$$
because $t-j+i\geq t.$  Then,  $\sum_{t=0}^\infty t^\zeta|\gamma_X(t)|<\infty$ and so, \textit{1.} in Lemma \ref{LL}  implies $T_1<\infty.$

Concerning $T_2,$ as $j> i$ and $t-j+i>0$, we can apply the $c_\zeta-$inequality (see \cite{Loeve} p.157) to
$t=(t-j+i)+(j-i)$ to obtain that there exists $c_\zeta > 0$ such that $t^\zeta\leq
c_\zeta(t-j+i)^\zeta+c_\zeta(j-i)^\zeta \leq 2c_\zeta(t-j+i)^\zeta.$ Thus,
$$T_2\leq 2c_\zeta\sum_{j=0}^\infty h_ja_j\sum_{i=0}^{j-1} h_ia_i \sum_{t=2j+1}^\infty
(t-j+i)^\zeta|\gamma_X(t-j+i)|\leq 2c_\zeta\sum_{j=0}^\infty
h_ja_j\sum_{i=0}^{j-1} h_ia_i \sum_{t=0}^\infty t^\zeta|\gamma_X(t)|.$$  Then, using the same tricks as for
$T_1$ we obtain that $T_2<\infty.$\\
For $T_3$, the fact that  $\sum_{t=0}^\infty t^\zeta|\gamma_X(t)|<\infty$, implies that   there
exists an $R>0$ such that $|\gamma_X(t)|\leq R$ for all $t\in\mathbb{Z}$. Therefore,
$$
T_3\leq R\left(\sum_{i=0}^\infty h_i a_i\right)\sum_{j=0}^\infty h_ja_j(2j)^\zeta (2j+1)
=: R\left(\sum_{i=0}^\infty h_i a_i\right) T_3^*.
$$
By {\it 1.} in Lemma \ref{LL}, to show that $T_3<\infty$, we only need to prove that $T_3^* < \infty$. Furthermore, applying Jensen inequality and {\it \ref{El}.}  in Proposition \ref{unop}, we have that
\begin{equation} \label{EqC.2}
\BBe[T_3^*]\leq \sum_{j=0}^\infty a_j^{1/2} (2j)^\zeta (2j+1) \alpha^{1/2} (1-  \alpha)^{j/2}.
\end{equation}
This last series is convergent ($\alpha \in (0,1)$). Hence, $T_3^*$ is finite almost surely and the proof is ended.
% The proof ends because, obviously, (\ref{EqC.2}) implies that $T_3^*$ is almost surely finite.
%\qed
\end{proof}

\subsection{Conditions to apply  Epps test}
In this subsection we analyze the theoretical behavior of Epps test when applied to the randomly projected process (see Theorem \ref{TheoConsist}). Moreover, in a corollary (Corollary \ref{CoroEpps}) we will show that if $\lambda$ is drawn randomly, then the  Epps test is consistent against many more alternatives.

Let us first state  Lemma \ref{cs} that gives  the consistency for the estimator of the spectral density function at zero defined in (\ref{fhat}).
Let us denote by $k_{lmno}(q,r,q+r;\la)$ the fourth-order cumulant of $Z_{0,l},$ $Z_{q,m},$ $Z_{r,n},$ and $Z_{q+r,o},$ where, for instance,  $Z_{q,m}$ is the $m$-th component of the vector $g(Y_q,\la)-g_{\mu_Y,\gamma_Y}(\la)$ (see Subsection \ref{SubsEpps}).

%In Lemma 2.2 of \cite{Epps} it is proved the  convergence $\hat f(0,\la)$ under the null hypothesis. However, thanks to the following Lemma we do not need to be under the null if we assume \ref{cumu}.
\begin{Lemm}\label{cs}
Let $\lambda \in \Lambda_N$.
If $\textbf{Y}$ is a stationary process such that
\begin{eqnarray}\label{cumu} \sup_{-\infty<q<\infty}\sum_{r=-\infty}^\infty |k_{lmno}(q,r,q+r;\la)|<\infty \mbox{ for each } l,m,n,o\in\{1,...,N\},\end{eqnarray}  then,
$\hat f(0,\la)\rightarrow f_{{\mathbf{Y}}}(0 ,(\mu_Y,\gamma_Y), \lambda) \mbox{ almost surely.}$
\end{Lemm}
\begin{proof}
\ \\
It is straightforward from the proof of Lemma 2.2 in \cite{Epps} but substituting by (\ref{cumu})  the use of (\ref{H}) and  Gebelein inequality, \cite{Gebelein}, for Gaussian processes. Gebelein inequality says that the autocovariance of a multidimensional process is smaller or equal than the product of variances of the marginals.
%\qed
\end{proof}

Lemma 3.1 in \cite{Epps} proves that if $\textbf{Y}$ is a stationary Gaussian process that satisfies (\ref{H}), then (\ref{cumu}) holds.
In \cite{GebeleinNuevo}, Gebelein inequality is extended to two-dimensional vectorial processes with diagonal densities. So that, any stationary process that satisfies (\ref{H}) and whose two-dimensional marginal has diagonal density, also satisfies (\ref{cumu}).

Let  $\Theta$ be an open and bounded subset of $\Rea\times\Rea^+.$
In \cite{Epps}, it is proved that \textbf{H1} and \textbf{H2} (see Subsection \ref{SubsEpps}) are satisfied if $\la_i$ is equal to a rational number times $\la_1,$ $i=2,...,N.$ Now, thanks to Lemma \ref{la} below, we have that $\la$ can be taken at random and still fulfill \textbf{H1} and \textbf{H2}.
\begin{Lemm}\label{la}
Assume that $\lambda =(\lambda_1  ,\ldots , \lambda_N)^T\in\Lambda_N$  ($N>1$) is drawn randomly and has distribution  $P_{\la}$ having the following properties.
First $P_{\la}$ is such that $\la_1$ and $\la_2$ are independent and identically distributed and have a density. Further, for $N>2,$ $\la_i$ is a rational number times $\la_1.$ Then, \textbf{H1} and \textbf{H2} are fulfilled almost surely.
%Let us take $\lambda =(\lambda_1  ,\ldots , \lambda_N)^T\in\Lambda_N,$ $N>1,$ at random using $P_{\la},$ where $P_{\la}$ is such that $\la_1$ and $\la_2$ are independent and identically distributed numbers chosen using a probability measure absolutely continuous with respect to the Lebesgue measure. In case $N>2,$ $\la_i$ is a rational number times $\la_1,$  $i=3,...,N.$ Then, \textbf{H1} and \textbf{H2} are fulfilled almost surely.
\end{Lemm}
\begin{proof}
\ \\
Proceeding as in \cite{Epps} we have that $$\Theta_0(\la)\subseteq \{(\nu,\gamma_Y): \nu\la_1=\mu_Y\la_1+2\pi k \mbox{ and } \nu\la_2=\mu_Y\la_2+2\pi k^*, \mbox{ with } k,k^*\in\mathbb{Z}\}.$$
Now, in order to get that the cardinal of $\Theta_0(\la)$ is larger than one, we need $\la_2$ is equal to a rational number times $\la_1.$
However, this happens with probability zero and so, with probability one $\Theta_0(\la)\subseteq\{(\mu_Y,\gamma_Y)\}.$ Thus,  \textbf{H1} and \textbf{H2} follow directly.
%\qed
\end{proof}
Note that in case $N>1$  Lemma \ref{la} remains valid if we draw independently at random $\la_i,$ $i=3,...,N.$
In addition, thanks to this lemma we have the following corollary of Theorem \ref{TheoEpps}.
\begin{Coro}\label{teo}
Let $(Y_t)_{t \in \mathbb{Z}}$  be a  stationary Gaussian process which satisfies (\ref{H}) and  $\la$ be as in Lemma \ref{la}. Let  $(\mu_n ,\gamma_n)$ be the minimizer on $\Theta$ of the map
$(\nu ,\rho) \conv Q_n(\nu ,\rho,\lambda)$ nearest to $(\hat\mu,\hat\gamma)$
 If we assume that $f_{{\mathbf{Y}}}(0 ,(\mu_Y,\gamma_Y), \lambda)$ is positive definite,
then $nQ_n(\mu_n ,\gamma_n,\lambda)$ converges in distribution to $\chi^2_{2N-2}.$
%a central chi-squared distribution with $2(N-1)$ degrees of freedom.
\end{Coro}

In the next theorem, the function $Q_n$ also depends on the random $\bf h$. However, for the sake of simplicity we have not express the functional relationship. %made explicit this dependence in order to keep the notation as simple as possible.

\begin{Theo} \label{TheoConsist}
Let \textbf X be an ergodic stationary process  satisfying (\ref{H}). Draw respectively $\la$ as in Lemma \ref{la} and $\textbf{h}$ independently of $\la$ using $P_{\bf H}$ (as described above).% and \textbf{Y} as .

 Assume that, conditionally on $\bf h$, $\bf Y$ defined in (\ref{Y}) satisfies (\ref{cumu}), that the characteristic function of its one-dimensional marginal is analytic and that $f_{{\mathbf{Y}} | \bf h}(0 ,(\mu_{Y |\bf h},\gamma_{Y |\bf h}), \lambda)$ exists and is positive definite for almost every $\bf h$.
 Let $Q_n(\cdot , \cdot,\lambda)$ be the quadratic form defined in (\ref{EqFCuadrat}) applied to \textbf{Y} and  $(\mu_n ,\gamma_n)$  its minimizer on $\Theta$ nearest to $(\hat\mu_{Y |\bf h},\hat\gamma_{Y |\bf h}).$ Let further $A:=\{(\la,h):nQ_n(\mu_n ,\gamma_n,\lambda)\rightarrow_d \mbox{ a non-degenerated distribution}\}.$

Then,   \textbf X is Gaussian  if, and only if,  $(P_{\la}\otimes P_{\bf H}) [A]>0.$

\end{Theo}
\begin{proof}
\ \\
Necessary part is obvious, because if \textbf X is Gaussian, then \textbf Y also is Gaussian and Proposition \ref{prop} %for  $\zeta>0$
implies that $\textbf{Y}$ satisfies the assumptions of Corollary \ref{teo}.\\
Let us show the sufficient part.
As $(P_{\la}\otimes P_{\bf H}) [A]>0$ we have that there exist ${\bf h}$ and $\la$ with $\la_1\neq 0$ and $\la_2\neq 0$  such that  $nQ_n(\mu_n ,\gamma_n,\lambda)$ converges in law to a non-degenerated distribution.  In addition, we may assume without loss of generality that $\Phi_{Y_0}(\la_1)\neq0$ and $\Phi_{Y_0}(\la_2)\neq0.$ Indeed, as $\Phi_{Y_0}$ is an analytic characteristic function it has only isolated zeros.\\
Therefore, $Q_n(\mu_n ,\gamma_n,\lambda)$ converges in probability to zero.
By Lemma \ref{cs},  $ \hat f(0,\lambda)$
 converges to $f_{{\mathbf{Y}}|\bf h}(0 , (\mu_{Y |\bf h},\gamma_{Y |\bf h}), \lambda).$ Thus, $\lim_nG^+_n$ is positive definite because it is the inverse of $2\pi f_{{\mathbf{Y}}|\bf h}(0,(\mu_{Y |\bf h},\gamma_{Y |\bf h}),\la)$. This together with (\ref{EqFCuadrat})  gives  that
\begin{eqnarray}\label{2g}\hat g(\lambda) -  g_{\mu_n ,\gamma_n}(\lambda)\convp 0. \end{eqnarray}
Since ${\mathbf{X}}$ is an ergodic stationary process, by \cite{Doob} page 458 we have that  $\left(g(Y_t,\lambda)\right)_{t \in \mathbb{Z}}$ is also an ergodic stationary process. Thus, as  $\BBe|\cos(\lambda_i Y_0)|<\infty$ and $\BBe|\sin(\lambda_i Y_0)|<\infty$ for all $i=1,...,N,$
we have by Theorem 2 of Chapter IV in \cite{Hannan} that $$\hat g(\lambda) \convp \BBe[g(Y_0,\lambda)].$$   From this and (\ref{2g}) we may conclude that  $\Phi_{\mu_n ,\gamma_n}(\lambda_i)$ converges in probability to $\Phi_{Y_0}(\lambda_i)$ ($i=1,...,N$).\\
Let us see how this implies  that the sequence $\{(\mu_n ,\gamma_n)\}_{n\in \mathbb N}$ converges.
We have that
\[
\lim_{n\rightarrow\infty}|\Phi_{\mu_n ,\gamma_n}(\la_1) |
=
\lim_{n\rightarrow\infty} e^{-\la_1^2\gamma_n/2}
=
| \Phi_{Y_0}(\la_1)|, \mbox{ in probability},
\]
and, since $\la_1\neq 0$ and $\Phi_{Y_0}(\la_1) \neq 0$, this implies that there exists $s \in \Rea$ such that $s=\lim_{n\rightarrow\infty}\gamma_n$ in probability.
Note that there exists $\theta\in[0,2\pi)$ such that $$\Phi_{Y_0}(\la_1)=|\Phi_{Y_0}(\la_1)| \exp(i \theta).$$ As $\la_1\neq 0$, if we take $m:=\theta/\la_1$, then, we have that
$\Phi_{Y_0}(\la_1)=\Phi_{m,s}(\la_1).$\\
Analogously, we have that
$| \Phi_{Y_0}(\la_2)|
=
\lim_{n\rightarrow\infty} e^{-\la_2^2\gamma_n/2},$
 in probability,
and as $s=\lim_{n\rightarrow\infty}\gamma_n$ we obtain \begin{eqnarray}\label{ese}|\Phi_{Y_0}(\la_2)|=e^{-\la_2^2s/2}.\end{eqnarray}
Denoting $r=\la_2/\la_1$,
we obtain that
\[
\frac{\Phi_{Y_0}(\la_2)}{|\Phi_{Y_0}(\la_2)|}
= \lim_n e^{ir\la_1 \mu_n}
=\left(\frac{\Phi_{Y_0}(\la_1)}{|\Phi_{Y_0}(\la_1)|}\right)^{r}= e^{ir\la_1m}.
\]
Together with (\ref{ese}) this gives $\Phi_{Y_0}(\la_2)=\Phi_{m,s}(\la_2).$

As $\lambda_2$ was drawn independently from $\lambda_1$ with a distribution absolutely continuous with respect to the Lebesgue measure and as $\Phi_{Y_0}$ is analytic, by Proposition \ref{Theo1Dimens} we get that $Y_0$ is Gaussian. Then, by Theorem \ref{TheoGausianas}, we obtain that the process  $\textbf{X}$ is Gaussian.
%\qed
\end{proof}
\begin{Nota}
{\rm It is only necessary to assume that \textbf{X} is ergodic to prove the inverse part in Theorem \ref{TheoConsist} since every stationary Gaussian process which satisfies (\ref{H})  is ergodic.
(see for example \cite{IbraRo})} %ver 26/3/09(2)
\end{Nota}

Applying the arguments of Theorem \ref{TheoConsist} directly to the process $\bf X$, we obtain the following corollary. It gives a modification of Epps test with better consistency properties.

\begin{Coro}\label{CoroEpps}
Let \textbf X be an ergodic stationary process. Assume that the characteristic function of its one-dimensional marginal is analytic. Assume further that (\ref{H}) holds. Let us take $\la$ as in Lemma \ref{la}, $Q_n(\cdot , \cdot,\lambda)$ as in (\ref{EqFCuadrat}), let $(\mu_n ,\gamma_n)$ be its minimizer on $\Theta$ nearest to $(\hat\mu_X,\hat\gamma_X)$ and $$B:=\{\la:nQ_n(\mu_n ,\gamma_n,\lambda)\rightarrow_d \mbox{ a non-degenerated distribution}\}.$$

If we assume that $f_{{\mathbf{X}}}(0 ,(\mu_X,\gamma_X), \lambda)$ exists and is positive definite, then, \textbf X is Gaussian if, and only if, $P_{\la} (B)>0.$
\end{Coro}

Remark below can be obviously deduced from Theorems \ref{TheoEpps} and \ref{TheoConsist}. This remark allows to perform a test based on the asymptotic distribution of $nQ_n(\mu_n ,\gamma_n,\lambda).$
\begin{Nota}\label{c}
{\rm Theorem \ref{TheoConsist} and Corollary \ref{CoroEpps} remain valid if we change in the definition of sets $A$ and $B$ ``non-degenerated distribution" by ``chi-squared distribution with $2(N-1)$ degrees of freedom".}
\end{Nota}
In addition, we have the following corollary.
\begin{Coro}
Under the assumptions of Theorem \ref{TheoConsist},  $(P_{\la}\otimes P_{\bf H}) [A]\in\{0,1\}$ and \textbf X is Gaussian if, and only if, $(P_{\la}\otimes P_{\bf H}) [A]=1.$

Analogously, under the assumptions of  Corollary \ref{CoroEpps},  $P_{\la} (B)\in\{0,1\}$ and
 \textbf X is Gaussian if, and only if, $P_{\la} (B)=1.$
\end{Coro}

\subsection{Conditions to apply Lobato and Velasco test}\label{ConLV}
In this subsection we show that a slight modification of the statistic $\tilde{G}_Y$ satisfies Theorem \ref{LoVe} under different assumptions than the ones used in \cite{LoVe}.

 The test statistic is $$G_Y=\frac{n\hat{\mu}_3^2}{6|\hat{F}_3|}+\frac{n(\hat{\mu}_4-3\hat{\mu}_2^2)^2}{24|\hat{F}_4|}$$ with
$$\hat{F}_k=2\sum_{t=1}^{\tau_n}\hat{\gamma}(t)(\hat{\gamma}(t)+\hat{\gamma}(\tau_n+1-t))^{k-1}+\hat{\gamma}^k,$$
where, according to Theorem \ref{teoKava}, we take $\tau_n<cn^{\beta_0}$ for $\beta_0=1-2/\alpha,$ $c>0$  and $2<\alpha<4.$ Thus, the differences between $G_Y$ and $\tilde{G}_Y$ are the absolute values in the denominator and the number of terms involved in $\hat F_k.$

\begin{Theo}\label{LoVenuestro}
Let $(X_t)_{t\in\mathbb{Z}}$ be an ergodic and stationary process such that  $\sum_{t=0}^\infty |\gamma_X(t)|<\infty.$ We have that
\begin{enumerate}
\item \label{1}
 If $(X_t)_{t\in\mathbb{Z}}$ is a Gaussian process, then $G_Y\longrightarrow_d \chi_2^2.$
\item  \label{2}
 If $(X_t-\mu_X)_{t\in\mathbb{Z}}$ can be written as (\ref{Kava}) and $\BBe[X_0^4]<\infty,$ then, conditionally on $\bf h$, $G_Y$ diverges almost surely to infinity whenever $\mu_3\neq 0$ or $\mu_4\neq 3\mu_2^2.$
\end{enumerate}
\end{Theo}
\begin{proof}
\ \\
Using Proposition \ref{prop} for $\zeta= 0$ we get that $(Y_t)_{t\in\mathbb{Z}}$ is an ergodic and  stationary process with $\sum_{t=0}^\infty |\gamma(t)|<\infty.$

If $(X_t)_{t\in\mathbb{Z}}$ is Gaussian, the process $(Y_t)_{t\in\mathbb{Z}}$ is also Gaussian. Thus, assumptions of the first part of Theorem \ref{LoVe} hold for the process $(Y_t)_{t\in\mathbb{Z}}$ and so $\tilde{G}_Y\longrightarrow_d \chi_2^2.$
Now, as \textbf{Y} is Gaussian, by \cite{Gasser} page 568, we have that $F_k>0$ for $k=3,4.$ Repeating the proof of Lemma 1 in \cite{LoVe}, we have that $\lim_{n\rightarrow\infty}\hat{F}_k=F_k$ and so, we may conclude that $\lim_{n\rightarrow\infty}G_Y=\lim_{n\rightarrow\infty}\tilde{G}_Y$ which shows {\it 1}.\\
Let us prove now   statement \textit{\ref{2}.} First, let us show that $\BBe[|Y|^k | {\bf h}]<\infty$, almost surely, for $k=1,...,4.$
By H\"older inequality we have that $|Y_0|\leq (\sum_{i=0}^\infty a_i)^{1/2}(\sum_{i=0}^\infty h_i^2a_i X_{-i}^2)^{1/2}$ and, as by Proposition \ref{unop} $\sum_{i=0}^\infty h_i^2a_i=1,$ almost surely,  we can apply Jensen  inequality. We obtain that
$$ Y_0^4\leq \left(\sum_{i=0}^\infty a_i\right)^{2}\left(\sum_{i=0}^\infty h_i^2a_i X_{-i}^4\right)
, \mbox{ almost surely}.$$
Thus,
$\BBe[|Y_0|^k | {\bf h}]<\infty,$ almost surely, for $k=1,...,4.$
By \cite{Doob}, page 458, we have that $\left( Y^k_t\right)_{t\in\mathbb{Z}}$ is stationary and ergodic, for all $k=1,...,4.$
Therefore, Theorem 2 of Chapter IV in \cite{Hannan} implies that
  \begin{eqnarray}\label{mu}
  \lim_{n\rightarrow\infty} \hat{\mu}_k=\mu_k, \mbox{ for almost every } {\bf h} \mbox{ and } k=2,3,4.
  \end{eqnarray}
Further, let us prove that $\lim_{n\rightarrow\infty}|\hat{F}_k|<\infty$ for almost every  {\bf h} and  $k=3,4.$
We have
\[
\hat {F}_k
=
\hat{\gamma}_Y^k+2\sum_{t=1}^{\tau_n}\sum_{j=0}^{k-1}\left(\begin{array}{c} k-1 \\ j\end{array}\right)\hat{\gamma}_Y^{}(t)^{k-j}\hat{\gamma}_Y^{}(\tau_n+1-t)^{j}.
\]
Taking into account that $|a^{k-j}b^j|\leq |a|^k+|b|^k,$ with $k\in\mathbb{N},$ $j\in\mathbb{N}$ and $j<k,$  we have
$$|\hat{F}_k|\leq  |\hat{\gamma}_Y^{}|^k+2^k\sum_{t=1}^{\tau_n}(|\hat{\gamma}_Y^{}(t)|^{k}+|\hat{\gamma}_Y^{}(\tau_n+1-t)|^{k}),$$
and then we obtain $|\hat{F}_k|\leq  2^{k+1}(\sum_{t=0}^{\tau_n}|\hat{\gamma}_Y^{}(t)|)^{k}.$
Let us prove now that
$$\lim_{n\rightarrow\infty}\sum_{t=0}^{\tau_n}|\hat{\gamma}_Y^{}(t)|<\infty.$$
Note that as $\BBe[X_0^4]<\infty,$ we also have
\begin{eqnarray*}
\infty>\BBe[(X_0-\mu_X)^4]
&=&\sum_{j_1,...,j_4=1}^\infty\prod_{r=1}^4k(j_r)E\left[\prod_{r=1}^4 \epsilon_{n-j_r}\right]
\\
&=&
\BBe[\epsilon_{1}^4]\sum_{j=1}^\infty k(j)^4+
\BBe[\epsilon_{1}^2]^2\sum_{i,j=1,  i\neq j}^\infty k(i)^2k(j)^2,
\end{eqnarray*}
because  $\left(\epsilon_{n}\right)$ are i.i.d.r.vs. with $\BBe[\epsilon_1]=0.$ So that
$\BBe[\epsilon_{1}^4]<\infty.$ Further, using Theorem \ref{teoKava} we obtain that
\[
\left|
\sum_{t=0}^{\tau_n} \left( |\hat{\gamma}_X(t)|-|\gamma_X(t)| \right)
\right|
\leq
 (\tau_n+1) o\left(n^{2/\alpha-1}\right)
=o(1).
\]
Thus, $\lim_{n\rightarrow\infty}\sum_{t=0}^{\tau_n}|\hat{\gamma}_X(t)|<\infty.$ Then,   by proceeding similarly as in the proof of Proposition \ref{prop}, we get $\lim_{n\rightarrow\infty}\sum_{t=0}^{\tau_n}|\hat{\gamma}(t)|<\infty$ and so, $\lim_{n\rightarrow\infty}|\hat{F}_k|<\infty$ for $k=3,4.$ Using (\ref{mu}) we may conclude that {\it 2.} holds.
%
%\qed
\end{proof}
Finally, applying Theorem \ref{LoVenuestro} directly to the process $\bf X$, we obtain the following corollary.
\begin{Coro}\label{LoVenuestroC}
Let $(X_t)_{t\in\mathbb{Z}}$ be an ergodic and stationary process such that $\sum_{t=0}^\infty |\gamma_X(t)|<\infty.$ We have that
\begin{enumerate}
\item \label{111}
 If $(X_t)_{t\in\mathbb{Z}}$ is a Gaussian process, then $G_X\longrightarrow_d \chi_2^2.$
\item  \label{22}
 If $(X_t-\mu_X)_{t\in\mathbb{Z}}$ can be written as (\ref{Kava}) and $\BBe[X_0^4]<\infty,$ then, conditionally on $\bf h$, $G_X$ diverges almost surely to infinity whenever $\mu_{X,3}\neq 0$ or $\mu_{X,4}\neq 3\mu_{X,2}^2.$
\end{enumerate}
\end{Coro}

\section{The tests in practice}\label{Simulations}

In this section we discuss the practical implementation of the gaussianity test. We start doing some remarks on  Epps test.

\subsection{Remark on Epps test}\label{subepps}
 Although Theorem \ref{TheoEpps} works for any $\lambda\in\Lambda_N,$ with $N>1,$ which satisfying \textbf{H1} and \textbf{H2}, in \cite{Epps} it is stated that:
{\it
\begin{itemize}
\item
When either $N$ is large or the spacing between the $\lambda_j$ is small, relative to the scale of the data, the matrix $2 \pi \hat f (0,\lambda)$ often appears
computationally singular.

\item
Also, values of $\lambda_j$ which are large, relative to the scale of the data, makes difficult to find a minimum of $Q_n(\cdot ,\cdot ,\lambda)$ with much precision.
\end{itemize}
}

Epps suggests to take \begin{eqnarray}\label{beta} \lambda_j=\xi_J/\sqrt{\hat\gamma}, \mbox{ with } \xi>0, j=1,...,N.\end{eqnarray}  Recall that $\hat\gamma$ denotes the sample variance of the process.
He proved  that  Theorem \ref{TheoEpps} works taking such $\lambda.$ In the simulations of Epps, and also in the ones of \cite{LoVe}, $N=2$ and $(\xi_1,\xi_2)=(1,2).$

 However, we need to draw $\la$ randomly in order to have a consistent test (Theorem \ref{TheoConsist}). So, we take $N=2,$ $\xi_1$ distributed as  the absolute value of a standard normal distribution and $\xi_2$ distributed  as the absolute value of a normal distribution with mean zero and variance $4.$ With this selection, although seldom, we have found that $\hat f(0,\la)$ could be singular. This is the main reason to choose $G^+_n(\lambda)$ as the generalized inverse of $2\pi\hat f(0,\la).$

Another important practical issue is  the procedure used to find the minimizer nearest to $(\hat\mu,\hat\gamma)$ of the map $(\nu ,\rho) \conv Q_n(\nu ,\rho,\lambda).$ In the simulations of \cite{Epps} and \cite{LoVe} they use the simplex method developed in \cite{amoeba}. We did the same. The code of such method can be found in \cite{NumericalRecipies}  under the name {\tt amoeba}.

\subsection{The random projection procedure to test Gaussianity} \label{SubsRPtest}
The theoretical development of Section \ref{sectionModelo} was carried out assuming that the observed sample is infinite. However, in practice, only a finite number of measurements $X_0,\ldots, X_n$ are available. So that, only a  finite number of  components of $\textbf{h}$ are computed.
This last difficulty is handled by fixing a small  $\delta>0$ (equal to $10^{-15}$ in the simulations that we present in Section \ref{simu}) and by  taking
$\textbf{h}=(h_0,\ldots, h_m)^T$ with
$$
m-1 =\min\{ \min\left\{t: \|(h_0,\ldots,h_t)^T\| \geq 1-\delta \right\},n-1\},
$$
where $h_0,\ldots, h_{m-1}$ are drawn by the stick breaking  procedure described in Section \ref{sectionModelo}. Further,  $h_m$ is fixed  such that $\|\textbf{h}\| =1.$
Concerning the projected process, some possibilities are available, but  here we use
\[
Y_t=\sum_{i=0}^{\min(m,t)} h_i X_{t-i}a_i, \mbox{ } t=0,\ldots, n.
\]

Let us give a short comment on the choice of the parameters  $\alpha_1,\;\alpha_2>0$ of the beta distribution used to generate $\textbf{h}.$  Here we have to deal with the following situation: If $m$ is large, then the random variables $Y_t $ are linear combinations of many random variables from the first sample and then, because of the Central Limit Theorem, the distribution of the random variables $Y_t $ will become closer to a normal law. That
%than to variables $X_t$;
will cause some loss of power when the marginal of {\bf X} is not Gaussian. Thus, in order to detect a non-Gaussian marginal, it is wise to  select $\alpha_1$ and $\alpha_2$  in such a way that $m$ is small or even $0$ or $1$.   This goal  is achieved if we take $\alpha_2=1$ and $\alpha_1 \gg 1$. Our selection in Section \ref{simu} is $\alpha_1=100$.
However, in this case the samples $Y_0,\ldots,Y_{n} $  and $X_0,\ldots,X_{n} $ are quite similar. So that, the test will not be good in detecting non-Gaussian alternatives with Gaussian marginal. In order to overcome this problem we should take $\textbf{h}$ in such a way that the projections mix several variables from the initial sample. To achieve this goal we need to take $\alpha_2 > \alpha_1$ but with  $\alpha_2 $ being not too big  to avoid  the effect of the Central Limit Theorem. In this case, a selection like $\alpha_1=2$ and $\alpha_2=7$ seems appropriate.
Therefore, it seems that in a practical situation we should decide which alternative is more plausible and then, select the appropriate parameters. However, there is another possibility: select two projections (one with each pair of parameters) and apply Theorem \ref{TheoFDR} to mix the $p$-values. This is our proposal.

Finally, we need a Gaussianity test for the one dimensional marginal  of $(Y_0,\ldots,Y_{n})$. We have seen two such tests (which have some advantages and disadvantages  discussed in Section \ref{simu}) and we can also mix them. Having all these requirements in mind,  we propose the following procedure:

\begin{enumerate}
\item
Draw $\textbf{h}^{(1)}$ with the $\beta(100,1)$ distribution and apply Epps test to the projections to obtain the $p$-value $p^{(1)}$.

\item
Draw $\textbf{h}^{(2)}$ (independently of $\textbf{h}^{(1)}$) with the $\beta(100,1)$ distribution and apply Lobato and Velasco test to the projections to obtain the $p$-value $p^{(2)}$.

\item
Draw $\textbf{h}^{(3)}$ (independently of $\textbf{h}^{(1)}$ and $\textbf{h}^{(2)}$) with the $\beta(2,7)$ distribution and apply Epps test to the projections to obtain the $p$-value $p^{(3)}$.

\item
Draw $\textbf{h}^{(4)}$ (independently of $\textbf{h}^{(1)}$, $\textbf{h}^{(2)}$ and $\textbf{h}^{(3)}$) with the $\beta(2,7)$ distribution and apply Lobato and Velasco test to the projections to obtain the $p$-value $p^{(4)}$.

\item
Combine the  $p$-values $p^{(1)},\ldots, p^{(4)}$ using the procedure described in Section \ref{MT} to decide  the Gaussianity hypothesis at the level $\alpha$. Thus, ordering these four $p$-values such that $p_{(1)}\leq ... \leq p_{(4)}$ we obtain that the $p$-value of the random projection test is equal to $(25/3)\cdot\min_{i=1,...,4}p_{(i)}/i.$

\end{enumerate}

\section{Simulations}\label{simu}
In this section we study the  behavior  of the proposed procedure in different situations. We have used  the same distributions as in \cite{LoVe}, in order to perform comparisons. Further, we will study a situation where the process has Gaussian marginal but is not Gaussian (see Section \ref{non-gaussian}). In addition, in Subsection \ref{real} we apply the random projection test to real data.

 The authors of \cite{LoVe} study the case of an AR(1) process depending on a parameter $q$ defined by
 \begin{eqnarray}\label{AR}
 X_{t}=qX_{t-1}+\varepsilon_t,
 \end{eqnarray}
where $q\in \{-.9, -.5, 0, .5, .6, .7, .8,.9\},$ $t\in \mathbb Z$ and $\varepsilon_t$ are i.i.d.  random variables with distribution $D_\varepsilon$ which may be  any of the following ones:
\begin{itemize}
\item standard normal (N(0,1)),
\item standard log-normal (log N),
\item Student $t$ with 10 degrees of freedom, ($t_{10}$),
\item chi-squared with $1$ ($\chi_1^2$) and $10$ degrees of freedom ($\chi_{10}^2$),
\item uniform on $[0,1]$ ($U(0,1)$),
\item beta with parameters $(2,1)$ ($\beta(2,1)$).
\end{itemize}
To simulate the process, we generate a large number of independent realizations $\varepsilon_t, \ t=1,\ldots, M$ with distribution $D_\varepsilon$ and we take
\begin{itemize}
\item
$X_1=\varepsilon_1$

\item
     $X_t=qX_{t-1}+\varepsilon_t,$ $t=2,...,M$

\end{itemize}
It is obvious that if $q\neq 0$, this process is not stationary. For instance, $\mbox{Var}[X_t]=\mbox{Var}[\varepsilon_1](1-q^{2t})/(1-q^2) $ which is not constant  and, obviously, the differences increase with $|q|$. In order to alleviate this problem, we disposed a certain number, {\it past}, of observations. We have taken $past$ equal to $1000$ and $n=M-\mbox{\it past}$ equal to $100, 500,1000,$ which are the sample sizes handled in \cite{LoVe}.

We have performed 5000 simulations in each situation. In every run we have computed the $p$-values using the asymptotic distributions. This could have caused that sometimes the rejection rates under the null hypothesis are  far away from the nominal level   (mostly for the lowest sample size $n=100$) and that they decrease under some alternatives  with the sample size (mostly for high values of $|q|$).

There are some differences between our rates and those published in \cite{LoVe}. We think they could be due to the fact that the {\it past} taken in \cite{LoVe} is not large enough.
For example, in the case $n=100$, $q=.7$ and  $D_\varepsilon$ being  $\beta(2,1)$
we obtain a rejection rate of $.2214$ when using  Epps test while in \cite{LoVe} they obtain one of $.080,$ which is appreciatively worse.
As explained before, our simulations were made with {\it past}$=1000$, but from Table \ref{past} we see that $.080$ is a rejection rate reasonable for {\it past} $=0$ and that the rejection rates increase with {\it past}, approaching to the value we have obtained.
\begin{center}
    \begin{tabular}
    {r|cccc} {\it past} &0&1&2&10
    \\
    \hline
    {\it rejections} & .0750&.1378&.1998&.2210\end{tabular}
    \end{center}
\begin{Tabla} \label{past}
Rejection rates along 5,000 simulations for different {\it past}, with Epps test,  $n=100,$ $D_\varepsilon$ a $\beta(2,1)$ and $q=.7.$
\end{Tabla}
We have observed the same problem with Lobato and Velasco test, excepting that with this other test our rejection rates are lower than those reported in \cite{LoVe}.

%\vspace{-5cm}
\begin{figure}[htb]
\begin{center}
\includegraphics[width=15cm, height=11cm]{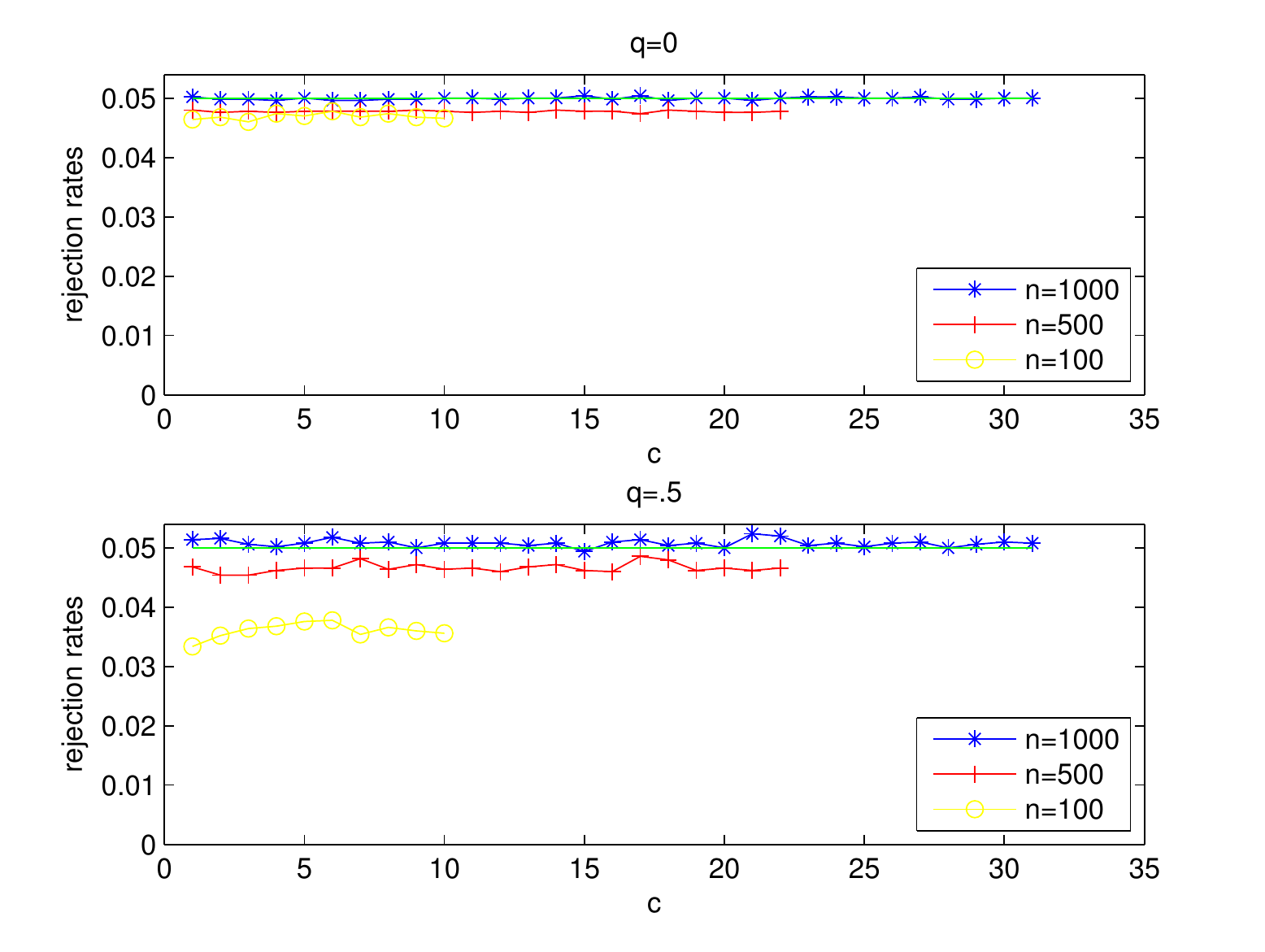}
\end{center}
%\vspace{-5cm}
\caption{Rejection rates under the null hypothesis for an AR(1) process with $q=0$ (upside graph), and  $q=.5$ (downside graph),  using Lobato and Velasco test, for different values of $c$ and sample sizes.}
\label{c5}
\end{figure}

Furthermore, another difference to bold %be taken into account
between what we do here and \cite{LoVe} is that in Subsection \ref{ConLV}  a sum until $\tau_n$ is involved in the estimation of $F_k$ while in  \cite{LoVe} the sum goes until $n-1,$ where $n$ is the sample size. Here, we have to take $\tau_n<cn^{\beta_0},$ where $\beta_0 =1-2/\alpha$ with $2<\alpha<4$  and $c>0$.
Thus, $\beta_0$ may be as close as desired to .5 and so,  we have decide to fix its value at $\beta_0=.5$ for the simulations.
In order to select the right value of $c$, we have made a small analysis to see how sensitive is the method to this parameter. We run Lobato and Velasco test under the null hypothesis for all values of $q$ and  $c=1,2,\ldots, c_n,$ where $c_n=\lfloor \sqrt{n} \rfloor$ and $n=100,500,1000.$  Therefore, $c_ {100}= 10,$ $c_ {500}= 22$ and $c_{1000}=31.$
The results suggest that the value of $c$ has not much influence in the rejection rates and so, we choose $c=1.$ The results for the cases $q=0$ and $q=.5$ appear in Figure \ref{c5}. It is worth saying that the situation for $q=-.9$ is a bit different than for the other values, as with  $q=-.9$ the rejection rates look constant till a point in which those rates strongly decrease.

Tables \ref{t1}, \ref{t2} and \ref{t3} contain the rejection rates for several procedures when applied at the level $.05$. Next we mention the procedures we have selected and make some comments on the results of our simulations.

\begin{enumerate}
\item \label{E}
\textbf{Epps test, E-test}. We take $(\xi_1,\xi_2)=(1,2)$ in (\ref{beta}).

It seems that this test behaves poorly when $D_\varepsilon$ is $t_{10}$. Moreover, broadly speaking,  its power decreases for the considered alternative distributions when $|q|$ increases, having low powers when $|q|=.9$. Note also that under the null hypothesis (excepting the case $q=0$ with $n=1000$), the rejection rates are above the level of the test and that they increase with $|q|.$

The power decreases when the sample size increases in the cases in which $|q|=.9$ and the alternative is $t_{10},$ $\chi_{10}^2,$ $U(0,1)$ or $\beta(2,1)$ (and even with $q=.8$ when $D_\varepsilon=t_{10}$).

\item \textbf{Lobato and Velasco test, G-test.}
The rejection rates displayed have been simulated using the statistic $G_X$. However, they are similar to those obtained using $\tilde G_X$.

 The {\bf G}-test has very low powers when $|q|$ is large, sometimes even lower than those of the {\bf E}-test. In addition it suffers from a lack of power when $D_\epsilon$ is $U(0,1)$ or $\beta(2,1)$. The rejections under the null hypothesis are above the level of the test only in 4 cases out of 24. In contrast with the {\bf E}-test, here the rejection rates under the null hypothesis decrease when $q$ increases.

\item  \textbf{Combined Epps and Lobato and Velasco test, GE-test.}
 In previous paragraphs we have commented some problems of the {\bf E} and {\bf G} tests which go, let us say, in opposite directions. In order to solve these problems we combine both tests, using  the multiple testing procedure presented in Section \ref{MT}.

As stated in  Subsection \ref{subepps}, the {\bf GE}-test  have been obtained by drawing  independently  $\xi_1$ with absolute value of a standard normal distribution and $\xi_2$ with absolute value of a normal distribution with mean zero and standard deviation  $2.$  However, it is worth noting that the rejection rates we have obtained have been a bit larger than in the case we take $(\xi_1,\xi_2)=(1,2).$

 We can observe from Tables \ref{t1}, \ref{t2} and \ref{t3} that this combination gives rejection rates between those of the {\bf E} and {\bf G}-tests although closest to the highest one, and, sometimes, even above. This is due to, as we have previously said, the rejection rates of {\bf E} are here a bit larger than when  $(\xi_1,\xi_2)=(1,2).$

\item \textbf{Random projection test, RP-test.}
We apply this test following the guidelines provided in Subsection \ref{SubsRPtest}.

When $q$ is negative and we are under the alternative, we always get the highest rejection rates with the {\bf RP}-test.
 The most striking behavior of this test happens for $q=.9$ and $D_\varepsilon = \chi^2_{10}$ and $\beta(2,1),$ where the {\bf RP}-test obtains rejection rates larger than $0.8$ while the second more successful test remains below $0.25$. For the remaining values, it happens that the rejection rates using the {\bf RP}-test are between the rates obtained with the {\bf E, G }and {\bf GE} tests but closer to the highest than to the lowest.

\end{enumerate}

\begin{center}
\begin{tabular}{rl|ccccccc}
$q$ & Test &N(0,1)&log N & $t_{10}$&$\chi_1^2$ & $\chi_{10}^2$ & $U(0,1)$ & $\beta(2,1)$
\\
\hline
&E& .1264   &.0508   &.1104   &.0656   &.1124   &.1390   &.1354
\\
-.9&G& .0292    &.1414    &.0310    &.0840    &.0332    &.0290    &.0266
\\
 & GE & .0942    &.1422    &.0908    &.1072    &.0920    &.1020    &.1010
\\
&RP& .1380    &.8070    &.1742    &.7576    &.3076    &.2620    &.3902
\\
\hline
&E& .0724   &.6780   &.0556   &.8514   &.2058   &.5408   &.4914
\\
-.5&G& .0504    &.9986    &.1692    &.9986    &.4602    &.0102    &.1696
\\
 & GE & .0774    &.9976    &.1582    &.9972    &.4552    &.4454    &.4154
\\
&RP& .0752    &.9998    &.1980    &1    &.5824    &.6404    &.7460
\\
\hline
&E& .0632   &.9616   &.0830   &.9964   &.5372   &.9918   &.9704
\\
0&G&.0458    &1    &.2820    &1    &.7898    &.5404    &.7520
\\
 & GE &.0732    &1    &.2402    &1    &.8074    &.8596    &.8706
\\
&RP&.0772    &1    &.2288    &1    &.7640    &.8496    &.9054
\\
\hline
&E& .0682   &.8594   &.0608   &.9582   &.2610   &.5618   &.5562
\\
.5&G&.0384    &.9990    &.1696    &.9982    &.4118    &.0010    &.1102
\\
 & GE &.0642    &.9990    &.1444    &.9988    &.4700    &.4680    &.4882
\\
&RP&.0750    &.9908    &.1132    &.9880    &.5226    &.3256    &.7500
\\
\hline
&E& .0710   &.6118   &.0582   &.8106   &.2006   &.3462   &.3650
\\
.6&G& .0358    &.9884    &.1162    &.9772    &.2858    &.0012    &.0592
\\
 & GE&.0640    &.9882    &.1144    &.9832    &.3218    &.2800    &.3086
\\
&RP    &.0802    &.9536    &.1030    &.9262    &.5164    &.2580    &.7744
\\
\hline
&E& .0838   &.3250   &.0626   &.4640   &.1492   &.2032   &.2214
\\
.7&G& .0260    &.9076    &.0814    &.8196    &.1610    &.0036    &.0334
\\
 & GE&.0714    &.9042    &.0866    &.8448    &.1998    &.1634    &.1802
\\
&RP   &.0784    &.8022    &.0926    &.7010    &.5754    &.2902    &.8060
\\
\hline
&E& .1034   &.1552   &.0810   &.2004   &.1324   &.1620   &.1596
\\
.8&G& .0206    &.6146    &.0466    &.4406    &.0708    &.0046    &.0166
\\
 & GE&.0726    &.6118    &.0796    &.4488    &.1122    &.1154    &.1136
\\
&RP&.0896    &.4928    &.0932    &.3264    &.6766    &.3950    &.8782
\\
\hline
&E& .1752   &.1264   &.1618   &.1368   &.1612   &.1870   &.1680
\\
.9&G& .0106    &.1558    &.0094    &.0714    &.0150    &.0054    &.0086
\\
 & GE&.1074    &.1844    &.0968    &.1190    &.0980    &.1182    &.1072
\\
&RP&.1168    &.1982    &.1174    &.1338    &.8702    &.6788    &.9662
\\
\hline
\end{tabular}
\end{center}
\begin{Tabla} \label{t1}
Rejection rates at level $.05$ of a process defined by (\ref{AR}). Sample size $n=100.$
\end{Tabla}
%\newpage

\begin{center}
\begin{tabular}{rl|ccccccc}
$q$ & Test &N(0,1)&log N & $t_{10}$&$\chi_1^2$ & $\chi_{10}^2$ & $U(0,1)$ & $\beta(2,1)$ %&$\Gamma(1,1)$&$\Gamma(1,2)$&Exp(10)%&Cauchy&$t_10$
\\
\hline
&E& .0744   &.3720   &.0584   &.2162   &.0712   &.0918   &.0850
\\
-.9&G& .0708    &.8838    &.0840    &.6202    &.1142    &.0462    &.0754
\\
 & GE&.0780    &.8604    &.0924    &.5400    &.1116    &.0866    &.0952
\\
&RP   &.0810    &.9990    &.2260    &.9928    &.6924    &.4630    &.6918
\\
\hline
&E& .0594   &1   &.1334   &1   &.7730   &.9924   &.9922
\\
-.5&G&.0472    &1    &.4580    &1    &.9960    &.9656    &.9976
\\
 & GE&.0476    &1    &.3784    &1    &.9912    &.9514    &.9914
\\
&RP   &.0490    &1    &.5090    &1    &.9998    &.9946    &1
\\
\hline
&E& .0566   &1   &.3292   &1   &.9982   &1   &1
\\
0&G&.0480    &1    &.7428    &1    &1    &1    &1
\\
 & GE&.0510    &1    &.6756    &1    &1    &1    &1
\\
&RP   &.0554    &1    &.6188    &1    &1    &1    &1
\\
\hline
&E& .0654   &1   &.1476   &1   &.8808   &.9918   &.9960
\\
.5&G& .0454    &1    &.4340    &1    &.9972    &.9704    &.9988
\\
 & GE & .0516    &1    &.3816    &1    &.9924    &.9504    &.9962
\\
&RP   &.0618    &1    &.2656    &1    &.9610    &.7440    &.9634
\\
\hline
&E& .0566   &.9998   &.1026   &1   &.7084   &.8286   &.9090
\\
.6&G& .0470    &1    &.3336    &1    &.9582    &.4678    &.8858
\\
 & GE&.0570    &1    &.2692    &1    &.9388    &.6944    &.8870
\\
&RP   &.0610    &1    &.1794    &1    &.8604    &.4730    &.9006
\\
\hline
&E& .0708   &.9996   &.0786   &1   &.4704   &.4042   &.5810
\\
.7&G& .0474    &1    &.1970    &1    &.7592    &.0644    &.4040
\\
 & GE&.0598    &1    &.1670    &1    &.7332    &.3640    &.5768
\\
&RP   &.0702    &1    &.1282    &1    &.6986    &.2616    &.8786
\\
\hline
&E& .0776   &.9780   &.0710   &.9638   &.2500   &.1948   &.2564
\\
.8&G&.0744    &.9998    &.0976    &.9980    &.3908    &.1524    &.2628
\\
 & GE&.0702    &.9998    &.1102    &.9978    &.3972    &.1848    &.2960
\\
&RP   &.0710    &.9986    &.0910    &.9908    &.6834    &.2484    &.9208
\\
\hline
&E& .1156   &.5708   &.0944   &.4674   &.1526   &.1430   &.1560
\\
.9&G& .0232    &.8356    &.0370    &.5404    &.0764    &.0138    &.0336
\\
 & GE&.0802    &.8708    &.0838    &.6378    &.1490    &.1092    &.1390
\\
&RP   &.0860    &.7996    &.0770    &.5510    &.8430    &.4818    &.9772
\\
\hline
\end{tabular}
\end{center}
\begin{Tabla} \label{t2}
Rejection rates at level $.05$ of a process defined by (\ref{AR}). Sample size $n=500.$
\end{Tabla}

\begin{center}
\begin{tabular}{rl|ccccccc}
$q$ & Test &N(0,1)&log N & $t_{10}$&$\chi_1^2$ & $\chi_{10}^2$ & $U(0,1)$ & $\beta(2,1)$ %&$\Gamma(1,1)$&$\Gamma(1,2)$&Exp(10)%&Cauchy&$t_10$
\\
\hline
&E&.0648   &.7836   &.0578   &.4572   &.0826   &.0888   &.0942
\\
-.9&G& .0902    &.9934    &.1206    &.8932    &.2448    &.0760    &.1358
\\
 & GE &.0880    &.9856    &.1002    &.8560    &.2190    &.1004    &.1450
\\
&RP   &.0940    &1    &.3344    &.9998    &.8686    &.5876    &.8056
\\
\hline
&E& .0530   &1   &.2574   &1   &.9764   &1   &1
\\
-.5&G& .0436    &1    &.6778    &1    &1    &1    &1
\\
 & GE&.0450    &1    &.6040    &1    &1    &1    &1
\\
&RP   &.0378    &1    &.7498    &1    &1    &1    &1
\\
\hline
&E& .0490   &1   &.5946   &1   &1   &1   &1
\\
0&G& .0546    &1    &.9364    &1    &1    &1    &1
\\
 & GE&.0486    &1    &.9162    &1    &1    &1    &1
\\
&RP   &.0422    &1    &.8734    &1    &1    &1    &1
\\
\hline
&E& .0550   &1   &.2534   &1   &.9966   &1   &1
\\
.5&G& .0482    &1    &.6788    &1    &1    &1    &1
\\
 & GE&.0424    &1    &.6016    &1    &1    &1    &1
\\
&RP   &.0484    &1    &.4348    &1    &.9994    &.9738    &.9996
\\
\hline
&E& .0566   &1   &.1718   &1   &.9580   &.9800   &.9974
\\
.6&G&.0472    &1    &.5112    &1    &.9996    &.9724    &.9996
\\
 & GE&.0464    &1    &.4234    &1    &.9996    &.9550    &.9986
\\
&RP   &.0584    &1    &.2812    &1    &.9902    &.7110    &.9804
\\
\hline
&E& .0594   &1   &.1162   &1   &.7720   &.6338   &.8632
\\
.7&G& .0418    &1    &.3104    &1    &.9744    &.3642    &.8830
\\
 & GE&.0558    &1    &.2380    &1    &.9672    &.5642    &.8724
\\
&RP   &.0598    &1    &.1754    &1    &.8888    &.3554    &.9036
\\
\hline
&E& .0690   &.9998   &.0720   &1   &.4342   &.2288   &.4108
\\
.8&G&.0500    &1    &.1638    &1    &.6804    &.0432    &.3284
\\
 & GE&.0670    &1    &.1294    &1    &.6708    &.2216    &.4450
\\
&RP   &.0654    &1    &.0996    &1    &.7144    &.1920    &.9076
\\
\hline
&E&.0902   &.9152   &.0880   &.7690   &.1836   &.1170   &.1686
\\
.9&G&.0346    &.9944    &.0636    &.9136    &.1574    &.0174    &.0574
\\
 & GE&.0690    &.9926    &.0798    &.9206    &.2178    &.1040    &.1596
\\
&RP&.0736    &.9844    &.0678    &.8580    &.8328    &.3946    &.9774
\\
\hline
\end{tabular}
\end{center}
\begin{Tabla} \label{t3}
Rejection rates at level $.05$ of a process defined by (\ref{AR}). Sample size $n=1000.$
\end{Tabla}

\subsection{A strictly stationary non-Gaussian process with Gaussian marginal}\label{non-gaussian}
In this subsection we discuss the behavior of the proposed procedure when used on  a non-Gaussian process with Gaussian marginal. We have worked with the process introduced in Example 2.3 in \cite{Cues91}. Its construction is explained here for the sake of completeness.

Let $p$ be a prime number, and let $Y_0$, $U$ and $\{Z_{m\cdot p}, m=0,1, \ldots \}$ be mutually independent random variables all uniformly distributed on $\{0,1,\ldots,p-1\}$.
Set
\begin{eqnarray*}
Z_{m\cdot p+k}& = & Z_{m\cdotp p} \oplus ( k  Y_0 )\ , \ k=0,\ldots, p-1, m=0,1,2,\ldots
\end{eqnarray*}
where $\oplus$ stands for sum modulus $p$. According to \cite{Cues91} the sequence $W_n = Z_{n+U}$ is composed by pairwise independent random variables and it is stationary. Moreover, these random variables are not mutually independent because, by construction, for every $m\in \Nat$ we have that
\[
Z_{m\cdot p}+ Z_{m\cdot p+1} + \ldots + Z_{m\cdot p+p-1} = p(p-1)/2,
\]
and so,
\begin{equation}\label{EqPi_1}
W_{mp-U} + W_{mp-U+1} + \ldots + W_{mp-U+p-1} = p(p-1)/2.
\end{equation}
Therefore, the knowledge of the random variables $W_{n-U} ,W_{n-U+1} , \ldots , W_{n-U+p-2} $ completely determines the value of $W_{n-U+p-1} $.

Now, given $k\in \{0,\ldots, p-1\}$, let $q_k$ be the quantile of order $k/p$ of the standard Gaussian distribution. For every $\nin$, let us  define the random variable  $W_n^*$  conditionally to $W_n$ as follows: If $W_n=k$, then draw the value of $W_n^*$ with a  standard Gaussian distribution conditioned to be in the interval $(q_k,q_{k+1})$, and independent of all the other random variables.

Since $W_n$ is uniformly distributed on $\{0,1,\ldots,p-1\}$, we obviously have that $W_n^*$ is a standard Gaussian r.v.. Moreover, the sequence $(W_n^*)$ inherits the remaining properties of $(W_n)$. It is a strictly stationary sequence of pairwise independent Gaussian random variables.

However, if $n>p-1$ and we are aware of the values $W_{n-U}^* , \ldots , W_{n-U+p-2}^*$, we can recover the values $W_{n-U} ,\ldots , W_{n-U+p-2} $ and, because of (\ref{EqPi_1}), we may deduce the value of $W_{n-U+p-1} $. With this information, we know to which interval $W^*_{n-U+p-1} $ belongs. Therefore, the random variables $(W_n^*)_n$ are not mutually independent and so, the process is not Gaussian.

We have simulated the previous process $5000$ times for different values of $p$ and sample sizes $n=100, 500, 1000.$  Then, we have applied the {\bf RP}  test at the level $\alpha=.05.$ The rejection rates appear in  Table \ref{NoGauss}.
\begin{center}
\begin{tabular}{r|ccccccc}
 &    $p=2$&    $p=3$ & $p=5$ & $p=7$&$p=11$&$p=13$&$p=17$
\\
\hline
$n=100$& .1448    &.1268    &.1676    &.1516    &.1602    &.1380    &.1146
\\
\hline
$n=500$ & .3698    &.3654    &.4938    &.5154    &.5822    &.5590    &.5588

\\
\hline
$n=1000$ & .6382    &.6386    &.6814    &.7250    &.7802    &.7608    &.7700
\\
\hline
\end{tabular}
\end{center}
\begin{Tabla} \label{NoGauss}
Rejection rates for different sample sizes applying the {\bf RP} test to the $\mathbf{W^*}$ process at the level $\alpha=.05$.
\end{Tabla}

For comparison, we show in Table \ref{Gauss} the rates of rejection when using the {\bf E}, {\bf G} and {\bf GE} tests in the case $p=5$.  Since these tests check for the non-Gaussianity of the marginal, the rejection rates are not too high. However, it is worth to pay some attention to the rejection rates in this table. To begin with, they are below the intended level (except GE with $n=100$), but, more surprisingly, they show some decrease when the sample size increases. We think that this is due to the fact that these tests see the process $\mathbb{W^*}$ as {\it more Gaussian  than a Gaussian process}.

The reason is that when we generate observations of a Gaussian process, {\it approximately} a proportion of $1/p$ observations are in the interval $(q_k,q_{k+1}),$  with $k\in \{0,\ldots, p-1\}.$ However, the process  $\mathbb{W^*}$ generate {\it exactly} a proportion of $1/p$ observations in each  interval $(q_k,q_{k+1})$. So that, it has a ``more Gaussian" behavior than expected. Consequently the rejection rates are lower than $.05$ and this fact becomes more apparent when $n$ increases.

\begin{center}
\begin{tabular}{r|ccc}
 &    $n=100$ & $n=500$ & $n=1000$
\\
\hline
E  &.0338 & .0266 &.0186
\\
\hline
G & .0372&.0336&.0326
\\
\hline
GE & .0520    &.0336    &.0206
\\
\hline
\end{tabular}
\end{center}
\begin{Tabla} \label{Gauss}
Rejection rates  using the E, G and GE tests of the $\mathbb{W^*}$ process with $p=5$, at the level $\alpha=.05$.
\end{Tabla}

\subsection{Real data}\label{real}
In this subsection we work with  the well-known Canadian lynx and Wolfer sunspot data in order to illustrate the behavior of the random projection test.
The Canadian lynx data consists on the annual record of the number of lynxes  trapped  in the Mackenzie River district of the North-West Canada for the period from 1821 to 1934 while the Wolfer sunspot data consists on the annual record of the sunspot activity in the period from 1700 to 1960. These data were used in \cite{Epps} and previously in \cite{subba}, obtaining in both cases that  the processes are not Gaussian.

We perform the random projection procedure to the lynx and sunspot data following the indications in Subsection \ref{SubsRPtest}. The obtained $p$-values are displayed in Table \ref{realdata} together with those gotten in \cite{Epps} and in \cite{subba}.
 \begin{center}
\begin{tabular}{r|ccc}
 &    RP & Epps & S.R. \& G
\\
\hline
lynx & $1.029\times 10^{-4}$ & $1.402\times 10^{-5}$ & $1.084\times 10^{-4}$
\\
\hline
sunspot & $1.314\times 10^{-6}$ & $7.356\times 10^{-6}$ & $2.818\times 10^{-4}$
\\
\hline
\end{tabular}
\end{center}
\begin{Tabla} \label{realdata}
$p$-values using the {\bf RP}-test and the  tests proposed in  \cite{Epps}  and in \cite{subba} for the lynx and sunspot data.
\end{Tabla}

In these examples we obtain $p$-values having approximatively the same magnitudes as those of  \cite{Epps} and  \cite{subba}.

\section{Discussion}
In this paper we have introduced the random  projection test, {\bf RP}-test, to check the Gaussianity of stationary processes. Given a sample, this test is based in a three steps procedure. First, it is required to draw a vector $\textbf{h}$ in a suitable Hilbert space. Then, the sample is projected on the one-dimensional space spanned by $\bf h$. Finally, we take advantage of the fact that, with probability one, the initial process is Gaussian if the marginal of the projected one is Gaussian. Therefore, we only need to use a test to check the Gaussianity of the marginal of  a stationary process. In the final step we use a combination of the Epps and Lobato and Velasco tests.

The comparison of the  {\bf RP}-procedure with the Epps and Lobato and Velasco tests (as well as with the combination of them) in situations in which the marginal is not Gaussian is not bad, and there are cases in which the proposed test is clearly better. Moreover, the {\bf RP} test is able to detect alternatives with Gaussian marginal, while   the other tests are not designed to do this task.

In spite of the rejection rates shown in Table \ref{NoGauss} are above the nominal level, they are not so high, mostly when the sample size is $100$. A simple way to improve these rates is to increase the number of  random projections using the correction described in Section \ref{MT}. From Table \ref{u} it can be seen how an increase in the number of employed random projections improves noticeably the rates.  In this table half of the projections are taken using the $\beta(100,1)$ distribution and the other half with the $\beta(2,7)$ and in each case we compute  half of the $p$-values with the {\bf E} test and the other half with the {\bf G} test.

\begin{center}
\begin{tabular}{r|cccc}
 &    $k=2$&    $k=3$ & $k=5$ & $k=8$
\\
\hline
$n=100$& .1448 & .1906    &.2288    &.2674
\\
\hline
$n=500$ & .3654 & .5772    &.6988    &.8064

\\
\hline
$n=1000$ &.6814 & .7688    &.8498    &.8628
\\
\hline
\end{tabular}
\end{center}
\begin{Tabla} \label{u}
Rejection rates for different sample sizes applying the {\bf RP} test with $2^k$ projections to the $\mathbb{W^*}$ process with $p=5$.
\end{Tabla}

\vspace{5mm}
 \noindent {\bf Acknowledgment.}
This research has been
carried out partially during a stay of  J.A. Cuesta-Albertos
at Universit\'e
Paul Sabatier, Toulouse, France, supported by the Spanish Ministerio de Educaci\'on under grant PR2009-0355.

 \end{document}